# Modeling Hydraulic Fracture Entering Stress Barrier: Theory and Practical Recommendations

Aleksandr Linkov [1], Liliana Rybarska-Rusinek [1], Ewa Rejwer-Kosińska [1,*]

[1] Rzeszow University of Technology, The Faculty of Mathematics and Applied Physics,
al. Powstancow Warszawy 12, 35-959 Rzeszow, Poland

**Abstract**

Numerical modeling of hydraulic fracturing is complicated when a fracture reaches a stress barrier. For high barriers, it may require changing of a computational scheme. Despite there are examples of modeling propagation through barriers, there is no general theory clarifying when and why conventional schemes may become inefficient, and how to overcome computational difficulties. The paper presents the theory and practical recommendations following from it.

We start from the definition of the barrier intensity, which exposes that the barrier strength may change from zero for contrast-free propagation to infinity for channelized propagation. The analysis reveals two types of computational difficulties caused by *spatial discretization*: (i) *general* arising for fine grids and aggravated by a barrier; and (ii) *specific*, caused entirely by a strong barrier.

The *asymptotic approach* which avoids spatial discretization is suggested. It is illustrated by solving *bench-mark* problems for barriers of arbitrary intensity. The analysis distinguishes three typical stages of the fracture penetration into a barrier, and provides theoretical values of the Nolte-Smith slope parameter and arrest time as functions of the barrier intensity.

Special analysis establishes the accuracy and bounds of the asymptotic approach. It appears that the approach provides physically significant and accurate results for fracture penetration into high, intermediate and even weak stress barriers.

On this basis, simple practical recommendations are given for modeling hydraulic fractures in rocks with stress barriers. The recommendations may be promptly implemented in any program using spatial discretization to model fracture propagation.

## 1 Introduction

Since early studies of hydraulic fracturing (HF) (e.g., Perkins and Kern 1961; Nordgren 1972), the impact of stress inhomogeneity on the fracture propagation has been taken into account. Specifically, channelized propagation along a pay-layer between impenetrable neighbors was simulated by the classical Perkins-Kern-Nordgren (PKN) model. However, with increasing pumping pressure, a fracture extends also into neighbor layers, what affects final footprints and openings, and so, the efficiency of a HF treatment. This stimulated effort to account for the fracture growth beyond a pay-layer, which resulted in successive development of pseudo-three dimensional (P3D) models (Settari and Cleary 1982, 1986; Palmer and Carroll 1982, 1983; Palmer and Craig 1984; Meyer 1986; Morales 1989; Warpinski and Smith 1989; Mack and Warpinski 2000; Adachi et al. 2007; Rahman and Rahman 2010; Dontsov and Peirce 2015; Linkov and Markov 2020). They complemented the PKN model by equations for out-of-layer growth.

All these models presume that the variation of in-situ stresses, caused by the difference in elastic modules of rock layers, impacts the fracture growth much more significantly than the difference in the modules itself. Then rock mass may be assumed homogeneous, while the assigned initial stresses change. Numerical results, obtained by solving problems for a crack, intersecting layers with different elastic modules and different assigned stresses (Erdogan and Biricikoglu 1973; Peirce and Siebrits 2001a), support the suggestion on the predominant impact of stress-contrasts. Other solutions for 3D layered structures with planar cracks (Lin and Keer 1989; Kuo and Keer 1995; Peirce and Siebrits 2001a, b; Markov 2020) agree with it, as well. This assumption notably simplifies computations by avoiding the need in building, storing and repeated using a special Green function for a layered media. It suggests an extension to general problems, when a crack arises and propagates in an inhomogeneous medium with initial (in particular, residual) stresses, when the sizes of inhomogeneities are comparable or exceed those of a considered crack. What concerns with inhomogeneities of sizes less than those of a crack considered, they may be accounted for through effective modules by well-developed methods (e.g., Nemat-Nasser and Hori 1993; Qu and Cherakaoui 2006). Thus the model of a homogenous medium with non-uniform initial stresses looks acceptable for solving practical problems, especially those like modeling HF, for which some of important input parameters (including rock structure, properties and in-situ stresses) are often uncertain or unknown and estimated quite roughly.

---

* Corresponding author: Ewa Rejwer-Kosińska. E-mail: e_rejwer@prz.edu.pl , ORCID: https://orcid.org/0000-0001-8788-2090
Contributing authors: linkoval@prz.edu.pl (Aleksandr Linkov); rybarska@prz.edu.pl (Liliana Rybarska-Rusinek).

For HF problems, the assumption of a homogeneous medium is additionally supported by the specific physical feature. The driving factor of a hydraulic fracture propagation is the *net*-pressure, which is the *difference* between the actual fluid pressure and a typical in-situ stress at a treatment depth. The net-pressure, being much less than the typical in-situ stresses, the spatial changes in the latter are of dominant significance. Thus in further discussion we focus on jump-like changes of in-situ confining stresses, while rock properties are assumed homogeneous. Furthermore, merely the case when the fracture enters the surface from the area with lower confining stresses will be discussed. In this case, the surface of the jump presents a stress *barrier*, which hampers the propagation and complicates numerical modeling.

The examples of truly 3D modeling propagation through a stress barrier by implicit, semi-implicit and stabilized explicit methods may be found in the papers by Peirce (2015, 2016), Zia and Lecampion (2019), Chen et al. (2020), Linkov, Rybarska-Rusinek and Rejwer-Kosińska (2023). Meanwhile, as to our knowledge, there are no special theoretical investigations peculiarities of hydraulic fracture penetration into a barrier.

Development of a theory, besides its insightful aspect, appears valuable for practical modeling of hydraulic fractures. Indeed, the examples given in the cited papers and our special calculations have shown that there are no computational difficulties if the propagation is not strongly channelized, so that the ratio length-to-height of the fracture footprint does not exceed 5. However, as has appeared from calculations for strongly channelized propagation, when the barrier is so strong that the length-to-height ratio exceeds 30, there arise complications (Chen et al. 2020, Fig. 19; Linkov, Rybarska-Rusinek and Rejwer-Kosińska 2023). The time expense drastically grows, exceeding first hours, and there appear clear signs of computational instability.

This paper aims to develop the theory of the fracture penetration into a barrier and, on this basis, to give practical recommendations to avoid complications in numerical modeling of hydraulic fractures in areas with stress barriers. It includes three parts. The first of them presents the theory and the asymptotic approach to model penetration into a barrier. The second contains verification of the asymptotic approach. The third offers practical recommendations for numerical modeling.

## 2 PART I. Theory of fracture penetration into stress barrier

Reproduce, for completeness and convenience, the conventional equations for planar propagation of a fracture in homogeneous rock (e.g., Adachi et al. 2007; Peirce and Detournay 2008; Peirce 2015; Linkov 2015, 2019; Chen et al. 2020). They include

the continuity equation

$$\frac{\partial w}{\partial t} = -div\boldsymbol{q} - q_l - Q = 0 \tag{1}$$

the Poiseuille-type equation

$$\boldsymbol{q} = -\left(\frac{w^{2n+1}}{\mu'}\right)^{1/n} |\boldsymbol{\nabla} p_f|^{1/n-1} \boldsymbol{\nabla} p_f \tag{2}$$

the elasticity equation

$$-\frac{E'}{8\pi}\int_S \frac{w(\xi)}{r^3}dS_\xi = p_f(\mathbf{x}) - \sigma_{0n}(\mathbf{x}) \tag{3}$$

Herein, $w$ is the opening, $t$ is the time, $\mathbf{q}$ is the fluid flux in the propagation plane, $q_l \geq 0$ and $Q$ are the terms accounting for, respectively, fluid leak-off and fluid influx into the fracture, $\mu' = 2[2(2n+1)/n]^n M$, $n$ is the fluid behavior index, $M$ is its consistency index (for a Newtonian fluid, $n = 1$, $\mu' = 12\mu$ with $M = \mu$ being the dynamic viscosity), $p_f$ is the fluid pressure, $E' = E/(1-\nu^2)$ is the plane-strain elasticity modulus, $E$ is the Young's modulus, $\nu$ is the Poisson's ratio, $r = \sqrt{(x_1 - \xi_1)^2 + (x_2 - \xi_2)^2}$, $\sigma_{0n}$ is in-situ traction normal to the fracture surface, the coordinates $x_1$ and $x_2$ of the right Cartesian system are located in the fracture plane, the coordinate $x_3$ is orthogonal to them; the normal to the plane has the direction of the $x_3$ axis. To simplify notation, compressive stresses and tractions, as well as the fluid pressure, are assumed positive (thus $\sigma_{0n} > 0$). The simplified forms of (1) - (3) for plain-strain and axisymmetric problems may be found in the papers by Adachi and Detournay (2002) and Savitski and Detournay (2002), respectively.

These equations are complemented with initial, boundary, fluid front propagation and fracture conditions. The *initial* condition commonly presumes zero opening along any perspective fracture surface before start of fluid pumping. When neglecting the lag between the fluid and fracture fronts, the condition of zero opening at each of front points $\mathbf{x}_*$ serves as the *boundary* condition $w(\mathbf{x}_*, t) = 0$. This condition is met identically when looking for a solution of the hypersingular equation (3) on the class of functions equal to zero at the fracture front. The fluid front *propagation* condition is expressed by the speed equation (e.g., Kemp 1990; Linkov 2015). Its formulation employs the asymptotic behavior of a solution, which, depending on a particular problem, may correspond to various propagation regimes (see, e.g., Spence and Sharp 1985; Desroches et al. 1994; Lenoach 1995; Garagash, Detournay and Adachi 2011; Linkov 2015). The *fracture* conditions define the very possibility and the direction of fracture propagation. Commonly, they are formulated in terms of linear elasticity fracture mechanics (LEFM), and the tensile mode (see, e.g., Rice 1968) is assumed. Then, the fracture conditions are:

$$K_I = K_{IC}, \quad K_{II} = 0, \quad K_{III} = 0 \tag{4}$$

where $K_I$, $K_{II}$, and $K_{III}$ are, respectively, the normal, shear plane strain and shear anti-plane stress intensity factors (SIFs); $K_{IC}$ is the critical SIF, defined by the strength of a material. The first of (4) defines the very possibility of a fracture growth, two remaining define the direction of the propagation. For a planar fracture, it is assumed that the third of equations (4) is met identically, while the second of (4) conventionally drops out from considerations, when the direction of propagation is assigned along the in-plane normal to a front curve. The discussion of a very complicated issue of criteria defining possibility and direction of *out of plane* propagation (e.g., Erdogan and Biricikoglu 1973) is beyond the objective of the present paper.

It is convenient to use the *net-pressure*

$$p_{net}(\mathbf{x}) = p_f(\mathbf{x}) - \sigma_0 \tag{5}$$

rather than the fluid pressure $p_f$ itself, by subtracting a reference rock pressure $\sigma_0$ from $p_f$. For certainty, $\sigma_0$ may be taken as the closing rock pressure near a borehole. Then defining the *stress contrast* as

$$\Delta\sigma(\mathbf{x}) = \sigma_{0n}(\mathbf{x}) - \sigma_0 \tag{6}$$

the elasticity equation may be written in terms of the net-pressure and stress contrast by changing $p_f$ to $p$, and $\sigma_{0n}$ to $\Delta\sigma$. Clearly, since $\sigma_0 = const$, we have $\text{grad} p_f = \text{grad} p$; hence in equation (2), the net-pressure $p_{net}$ (5) may replace the fluid pressure $p_f$, as well. These replacements are assumed in further discussion; from now on, merely net-pressure and stress contrast are considered.

### 2.1 Concept and range of barrier intensity

To start a theory, it is necessary first of all to distinctly define what is the barrier *strength*. In the Introduction it is mentioned that barrier is a surface of sudden increase of the in-situ stress hampering fracture propagation. Thus the positive stress-contrast (6) on the propagation path presents a *dimensional* characteristic of a stress barrier. Clearly with growing stress contrast, the strength of barrier increases.

However, it remains unclear which values of this *external* parameter refer to low, intermediate, or high barriers. Much depends also on the internal parameters driving the fracture growth. This may be seen from Fig. 1, illustrating the asymptotic fields near a fracture tip entering a barrier.

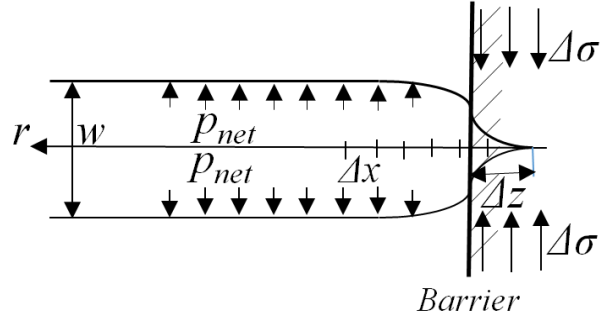

**Fig. 1** Asymptotic scheme for a fracture entering stress barrier

It shows that the internal quantity with the stress dimension is the average net-pressure $p_{av}$ near the crack tip. Thus an appropriate *non-dimensional* parameter, characterizing hydrofracture penetration, is the ratio

$$R = \frac{\Delta\sigma}{p_{av}} \qquad (7)$$

This parameter defines the actual strength of a stress contrast in reference to the net-pressure activating the penetration. We shall call this important parameter the barrier *intensity*. Since the net-pressure depends on the pumping rate, fluid viscosity, compliance of rock and it changes in time, the intensity $R$ strongly depends on these factors.

Of special significance is the *starting* value $R_0$ of intensity (7) at the moment, when the fracture front reaches the barrier. It may be evaluated by using the self-similar solutions to plane-strain (Fig. 2a) and axisymmetric (Fig. 2b) problems (Adachi and Detournay 2002; Savitski and Detournay 2002; Linkov 2015).

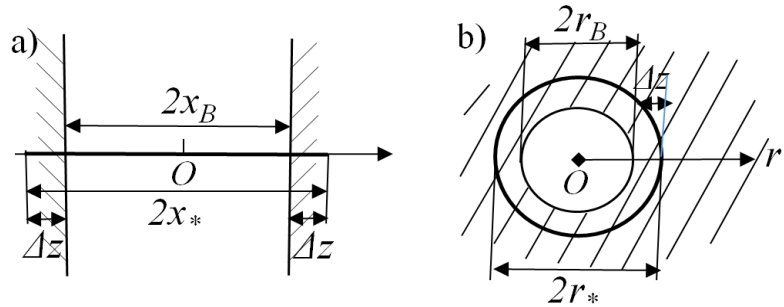

**Fig. 2** Schemes for plane-strain (a) and axisymmetric (b) problems of a fracture entering stress barrier

By using these solutions, it may be inferred that at the moment of reaching a barrier at the distance $H/2$ from the inlet, the barrier intensity for a Newtonian fluid is

$$R_0 = k_0 \frac{\Delta\sigma}{E'} \left(\frac{H^m E'}{Q_0 \mu'}\right)^{\frac{1}{4}} \qquad (8)$$

where $k_0 = 2.36$, $m = 2$, $H = x_B$ for the plain-strain problem (Gladkov and Linkov 2018), and $k_0 = 1.63$, $m = 3$, $H = r_B$ for the axisymmetric problem (Linkov, Rybarska-Rusinek and Rejwer-Kosińska 2023). From (8), it may be seen that the barrier intensity may change in a very wide range from zero for contrast-free propagation to large values, characterizing practically impenetrable barriers, for which $R_0 = \infty$. In particular, with decreasing pumping rate, the intensity $R_0$ goes to infinity, and a barrier with whatever *small* stress contrast, becomes *impenetrable*. In the opposite case, with increasing pumping rate, a barrier with whatever *large* stress contrast becomes *"invisible"* for fracture propagation. From the results of (Linkov, Rybarska-Rusinek and Rejwer-Kosińska 2023), it may be concluded that a barrier becomes practically impenetrable when $R_0$ exceeds 7; it is practically invisible when $R_0$ is less than 1. Below, these estimations are confirmed by special analysis.

With the intensity defined, the questions arise: if, when, why and how a barrier of a given intensity may cause computational complications? The fact that complications sometimes arise follows from the failure of both explicit and implicit algorithms established in the paper (Gladkov and Linkov 2018) for Khristianovich-Geertsma-de-Clerk (KGD) problem (Zheltov and Khristianovich 1955; Geertsma and De Klerk 1969) with stress contrasts. Evident signs of instability and extreme growth of time expense for calculations has been also observed in truly 3D modeling of fracture propagation within and out of a pay-layer between half-spaces, in cases when the fluid viscosity was small while the stress contrasts grew (Chen et al. 2020; Linkov, Rybarska-Rusinek and Rejwer-Kosińska 2023). As follows from the results of these papers, it happens for $R_0 > 3.5$, when the propagation becomes practically channelized.

The common reasons of these unfavorable computational effects consist of (i) unacceptably small Courant time *cou* for *explicit* Euler time stepping; (ii) growth of the condition number *cond* for *implicit* stepping with repeated matrix-to-vector multiplications, used for iterative solving a non-linear algebraic system on each step; (iii) enormous growth of the number of work units (*NWU*) for *stabilized explicit* and *implicit* schemes. These unfavorable factors are discussed in the next Subsection 2.2 with exposing the main reasons of their aggravation by a barrier. The discussion provides the utmost value of *cou*, below which the presently used schemes of solving the system (1) – (3) become unstable or/and inefficient.

Furthermore, in addition to these general causes of instability, there is a specific computational difficulty, caused entirely by a barrier when the pressure gradient near the barrier goes to zero. This specific reason of computational instability is explained in Subsection 2.3.

## 2.2 *General* restrictions of time stepping schemes aggravated by stress barrier

*Explicit Euler scheme.* The general computational difficulties of modeling hydraulic fracture are due to the inherent mathematical difficulty: after spatial discretization, the resulting system of ordinary differential equations (ODE) is *extremely stiff* (Peirce and Siebrits 2005; Peirce 2006). Its Courant time *cou* (e.g., Lebedev 1997) is commonly very small. The fundamental estimation for the *cou* is obtained by Peirce and Siebrits (2005). They studied the spectrum of the ODE main matrix for the plane-strain KGD problem with fixing the *average* value $w$ of the opening at the central part of a HF:

$$cou = t_n \left(\frac{\Delta x}{w}\right)^3 \tag{9}$$

where

$$t_n = \left(\frac{\mu'}{E'}\right)^{1/n} \tag{10}$$

The Courant time *cou* defines the maximal admissible time step $\Delta t_E$ for *explicit Euler* integration (e.g., Epperson 2011). According (9)

$$\Delta t_E = cou = t_n \left(\frac{\Delta x}{w}\right)^3 \tag{11}$$

Detailed calculations with using the explicit Euler scheme for the case of the benchmark plane-strain KGD problem, provided the data to check the accuracy of (11) (Linkov 2019, p. 16, 28, Fig. 4). It appears that (11) perfectly agrees with the bounds of stability found in numerical experiments. Specifically, the numerical results reproduce the cubic dependence of *cou* on the ratio $\frac{\Delta x}{w}$ and the linear dependence on the intrinsic time $t_n$, defined in (10). The factor (actually, 1.00) in front of $t_n$ also complies with the numerical value to the accuracy of 1%.

A non-linear system of ODE, arising after spatial discretization of a the system (1)-(3), involves a non-symmetric fully populated square matrix. In 3D problems, its order $N_{mat}$ often exceeds first hundreds. This implies that on each time step of numerical integration of the ODE, the corresponding algebraic system is to be solved iteratively with performing *matrix-to-vector* multiplications. The time-costly operation of matrix-to-vector multiplication, which requires $N_{mat}^2$ *arithmetic* multiplications, presents the work unit ($WU$) of numerical modeling.

For the *explicit Euler scheme*, each time step consists of a *single WU*. This allows us to easily estimate the total number of work units ($NWU$), computational complexity and the computer time expense for tracing fracture propagation on a typical time interval $T$ of a hydraulic treatment. Specifically, by (11), for the explicit Euler scheme, the number of work units is

$$NWU_E = \frac{T}{\Delta t_E} = \frac{T}{t_n}\left(\frac{w}{\Delta x}\right)^3 \qquad (12)$$

With known $NWU$, computational complexity $C$ and the times expense $T_C$ for tracing HF become known as well: $C = N_{mat}^2 NWU$, $T_C = \Delta t_a C$, where $\Delta t_a$ is the processor time for a single arithmetic multiplication with an assigned precision (for a conventional laptop, $\Delta t_a \approx 0.5 \cdot 10^{-8}$ s for ordinary, and $\Delta t_a \approx 1.0 \cdot 10^{-8}$ s for double precision arithmetic).

Exemplify these estimations by taking typical values $t_n = 10^{-11}$ s, $\Delta x = 2.5$ m, $w = 2 \cdot 10^{-3}$ m, $T = 3600$ s, $N_{mat} = 200$, $\Delta t_a = 10^{-8}$ s. Then $cou = \Delta t_E = 0.020$ s, $NWU_E = 184320$, $C_E = 7.4 \cdot 10^9$, $T_{CE} = 74$ s. This time is acceptable for numerical modeling with explicit Euler stepping.

However, for modeling *propagation through a barrier*, the need in finer grids arises (e.g., Peirce and Siebrits 2005). When taking an order finer grid and considering an order thinner fracturing fluid, the Courant time $cou$ becomes four orders less ($cou = 2 \cdot 10^{-6}$ s), while the number of work units $NWU_E$, complexity $C_E$, and time expense $T_{CE}$ become four orders greater. Thus the time cost $T_{CE} = 7.4 \cdot 10^5$ s for calculations on a conventional laptop reaches hundreds of hours. The explicit Euler time stepping becomes prohibitively expensive.

Then there arises the need in using either *implicit* (e.g., Epperson 2011) or *stabilized explicit* (e.g., Lebedev 1997; Meyer, Balsara and Aslam 2014) schemes. Consider the limitations of these methods when applied to penetration into a barrier. We start from the stabilized explicit schemes as natural improvement of the classical explicit Euler stepping.

***Stabilized explicit Runge-Kutta schemes.*** These schemes where developed specially to smooth the quite unfavorable limitation of the Courant-Friedrich-Levi (CFL) condition of instability (Lebedev 1997; Meyer, Balsara and Aslam 2014). The explicit stabilized Runge-Kutta-Legendre 2$^{nd}$ order (RKL2) method, suggested and studied by Meyer, Balsara and Aslam (2014), has appeared to be highly stable, accurate and efficient. Its first applications to hydraulic fracture modeling (Chen et al. 2020) has evidently demonstrated its advantages over other presently used methods employing spatial discretization.

The RKL2 method affords easy estimation of the corresponding number of work units $NWU_{RKL2}$. Denote $N_S$ the (odd) number of stages, employed on a super step of the RKL2 method (Meyer, Balsara and Aslam 2014). The required number of work units performed on a super-step is

$$NWU_{RKL2} \approx \frac{4}{N_S} NWU_E \qquad (13)$$

Equation (13) shows that the efficiency of the RKL2 method is $N_S/4$ - fold greater than the efficiency (12) of the explicit Euler scheme. In practical calculations for HF modeling by the RKL2 method the number of stages $N_S$ does not exceed 399 (in average, it is on the level of 111). Further increasing $N_S$ leads either to inacceptable growth of the time step, so that the front propagates several grid cells, or, what is worse, to notable influence of rounding errors. With the maximal $N_S = 399$, the efficiency (13) of the RKL2 method is 100-fold greater than of the explicit Euler.

Thus the number of work units $NWU_{RKL2}$, computational complexity $C_{RKL2}$ and time cost $T_{CRKL2}$ of the most advanced modern integration scheme for tracing HF are 100-fold less than those given above for explicit Euler scheme. When employing the RKL2 method, the critical value $cou_{RKL2}$ of the Courant time (9), acceptable for practical calculations, may be taken 100-fold less than its value acceptable when using explicit Euler. We see that the Courant time $cou$ is also of prime significance for the *time expense* of a HF modeling by the *explicit stabilized* Runge-Kutta (RK) methods.

The numerical data, obtained by the RKL2 method for the HF problem with parallel barriers, give an estimation of the minimal $cou$, which is close to the bound, beyond which the time expense of the modeling becomes enormous. From the results of modeling channelized propagation of a fracture driven by water with the dynamic viscosity $\mu = 1\, mPa \cdot s$ (Chen et al. 2000; Linkov, Rybarska-Rusinek and Rejwer-Kosińska 2023), it appears that tracing the propagation time of duration $10\, min$ required the computation time expense of 3 hours. The corresponding Courant time, calculated by using (9), is $cou = 1.75 \cdot 10^{-4}\, s$. To model further propagation up to the duration of 1.5 hour, would require much greater time expense. Roughly, the less restrictive critical $cou_{cr} = 1.5 \cdot 10^{-4}\, s$ may be taken as an acceptable bound. Thus to model 3D propagation of a HF on a given spatial grid, the following condition of computational efficiency is to be met:

$$t_n \left(\frac{\Delta x}{w}\right)^3 > cou_{cr} \tag{14}$$

where $cou_{cr} \approx 1.5 \cdot 10^{-4}\, s$. The condition (14) may serve to control if numerical modeling with spatial discretization is possible with using the quite advanced explicit stabilized RKL2 method. Our calculations by this method for channelized fracture propagation between *high barriers* ($R_0 > 10$) show that when the driving fluid is thin, in particular water ($\mu = 1\, mPa \cdot s$), the modeling is nearly on the limit of computational potential. Notably, the condition (14), like (9) and (12) distinctly show rapid growth of computational complications with the growth of the opening $w$ at near-barrier zone and/or with decreasing the grid size $\Delta x$.

***Comment on implicit schemes.*** Employing *implicit* schemes is another option for overcoming limitations of the explicit Euler integration. For them, the high stiffness of the system appears through high condition number ($cond$) of the matrix. Since implicit integration allows time step $\Delta t$ much greater than that of the Euler step (11), we may employ the analysis of the spectrum given in (Peirce and Siebrits 2005; Peirce 2006) to conclude on the typical condition number. Omitting details, the result is

$$cond \approx 2 \frac{\Delta t}{cou} \tag{15}$$

From (15) it appears that again the Courant time $cou$ is of prime significance for both the stability and efficiency of computations. Using (9) in (15) shows that the condition number rapidly, as $(w/\Delta x)^3$, grows with decreasing grid size $\Delta x$ and growing average opening $w$. This implies (Peirce and Siebrits 2005; Peirce 2006) that to avoid enormous number of iterations in the internal cycle, it is desirable to develop a proper *preconditioner*. The need in an effective preconditioner further increases when there are *jumps of in-situ* confining *stress*. For such problems, two efficient preconditioners have been specially designed in (Peirce and Siebrits 2005; Peirce 2006). Both of them distinguish the highly oscillating part of the matrix spectrum by "discarding all by nearest-neighbor influences" (Peirce and Siebrits 2005, p. 1807). Besides, the first of the preconditioners employs the *multigrid* approach (e.g., Briggs, Henson and McCormick 2000) to detect the input of high frequencies, generated by the area near a stress jump. The authors used two grids, so that the fine grid had the size two-fold less than the rough. Thus the fine grid could catch perturbations, caused by the jump, in an area of its size. The second of the preconditioners (Peirce 2006) did not directly account for the local perturbation; rather, to decrease the number of iterations, it employed approximate inversion of the distinguished part of the matrix. Their superiority over conventional preconditioners was demonstrated in the cited papers for an example when a fracture enters a layer with *accelerating* (negative) stress contrast.

Recall however, that an implicit method requires inversions of the matrix and repeated matrix-to-vector multiplications within a time step. This burden is quite unfavorable for computational complexity, and consequently for the time expense. As established in the paper by Chen et al. (2020, p. 370), modeling HF by the *stabilized* explicit RKL2 method "can be up to 30 times faster" as compared to the *implicit* integration. Hence, the limitations on the computational efficiency, revealed for the RKL2 method, certainly refer to explicit methods, as well. In particular, excessive time expense, indicated for the channelized propagation between *barriers of high intensity*, is even greater when employing implicit methods.

The considerations above show that the options for modeling a HF with spatial discretization are restricted even in the case of *parallel barriers*, when there is a passageway along which a fluid may flow and, consequently, the fracture may propagate. The *general* computational difficulties discussed further increase when the passageway itself is blocked by high stress barriers. Moreover, there is a *specific* difficulty, exposed in the next subsection, due to which any method of modeling with spatial discretization may become inapplicable.

**2.3** *Specific* **computational difficulty for modeling high barriers**

The core physical reason of the difficulties is revealed by considering the extreme case, when at the moment $t_B$ the *entire* fracture contour has reached a very high, in limit impenetrable, stress barrier. Then at $t = t_B$, the propagation speed $v_*(t_B - 0) = v_{*B} \neq 0$ instantly jumps to zero ($v_*(t_B + 0) = 0$). This drastically changes the asymptotics of fields near the fracture contour. Now the speed is zero ($v_*(t) \equiv 0$) for any $t > t_B$. Hence, the fluid particle velocity is also zero at the barrier. Then in a vicinity of the barrier, the key suggestion, accepted when deriving the Poiseuille-type equation, that in-plane component of the particle velocity is much greater than the component normal to the channel walls, is invalid. The Poiseuille-type equation becomes inapplicable. Physically, a reverse flow, caused by reflection of fluid from the impenetrable barrier, arises. Its detailed description requires inclusion into an analysis the fluid compressibility, the corresponding inertial term and, at least near the barrier, cross-sectional component of the particle velocity. This tremendously complicates numerical solution of the problem (see e.g., Cao, Hessein and Schreffler 2018).

However, with the time growth, an approximate limiting picture for a viscous fluid tends to become quite simple because of damping influence of viscous losses. As usual, due to viscosity, the flow "forgets" about a state fairly before the current time (cf., e.g., Linkov 2016b, c). With the time growth, the pressure between impenetrable barriers tends to become uniform. For a uniform net-pressure and unchanged fracture contour, from the elasticity equation it follows that the opening between the barriers changes proportionally to the net-pressure with unchanged form of the distribution. Hence, with the growth of time, the average opening and the uniform net-pressure become connected by a factor depending merely on the time. The global mass conservation implies that the average opening and, consequently the net-pressure, grow proportionally to the fluid volume between the barriers. For a constant pumping rate, the growth is linear.

These specific features of the case considered have unfavorable computational consequences for conventional evaluation of the fluid flux $\boldsymbol{q}$ by using the Poiseuille-type equation (2). The latter implies that $|\boldsymbol{q}|^n = \frac{w^{2n+1}}{\mu'}|\text{grad}p|$, so that for a finite flux, the pressure gradient decreases with time, at least as $1/w^{2n+1}$. Hence the conventional calculation of the flux involves the product of the term, which grows in time as $w^{2n+1}$, by the term, which decreases as $1/w^{2n+1}$. With growing time, conventional calculations unavoidably involve the uncertainty of the type $\infty \cdot 0$. In particular, for a Newtonian fluid ($n = 1$) and constant pumping rate ($Q(t) = Q_0$), commonly used in calculations, the terms are of orders $t^3$ and $1/t^3$, respectively. When the time changes three orders from an initial to a final value, each of the factors changes 9 orders with the second of them going to zero. Clearly, practical computations, even performed with arithmetic of double precision, would deteriorate when using values of openings in grid cells. The deterioration is due to the uncertainty.

We see that the deterioration of a conventional scheme will occur for any grid size and for any time step. It is caused by the specific features of the particular problem. The error blows up due to the uncertainty $\infty \cdot 0$ when calculating the flux by using the Poiseuille-type equation.

Note, however, that in essence the difficulties discussed, both general and specific, arise as a consequence of *spatial discretization*. Indeed, the influence of the latter appears through the *cubed grid size* $\Delta x^3$ in equations (9), (12), (14), and, by (9), also in (15), which define the stability and efficiency of modeling, and through the *nodal* values of *cubed opening* $w^3$ (for a Newtonian fluid) in the Poiseuille equation. This suggests a means to overcome the difficulties by avoiding spatial discretization and explicit use the Poiseuille equation. It is reached by the asymptotic approach.

**2.4 Asymptotic approach**

**2.4.1 Asymptotic solution in general case**

***Problem formulation in terms of SIFs.*** The conclusion on the jump to uniform pressure has far-reaching implications for our theme. It suggests using the approximation of *uniform pressure* in the cross-section orthogonal to a barrier. Then, as follows from the elasticity theory (e.g., Muskhelishvili 1975), the dominating asymptotics near the barrier are of the square-root type. The factor, defining their intensity, is the classical tensile stress intensity factor (SIF) (e.g., Rice 1968). Then employing SIFs becomes the natural means to study penetration into a barrier. For years, the suggestion on the uniform cross-sectional net-pressure between plane parallel barriers and the option to use SIFs have been employed for pseudo-three dimensional (P3D) modeling (e.g., Settari and Cleary 1982, 1986; Palmer and Carroll 1982, 1983; Palmer and Craig 1984; Meyer 1986; Morales 1989; Warpinski and Smith 1989; Mack and Warpinski 2000; Adachi et al. 2007; Rahman and Rahman 2010; Dontsov and Peirce 2015; Linkov and Markov 2020). The differences between various P3D models aroused merely when accounting for the viscous losses by means of the apparent viscous SIF $K_{IA}$. Actually, the proper way to define $K_{IA}$ for *symmetric* stress-contrasts has been found by Dontsov and Peirce (2015). Its general form for *arbitrary* contrasts is expressed by the correspondence principle (Linkov and Markov 2020). The applicability of this principle in a wide range of stress contrasts has been confirmed by good agreement of the foot prints obtained in comparative P3D and truly 3D calculations (Dontsov and Peirce 2015; Linkov and Markov 2020; Linkov, Rybarska-Rusinek and Rejwer-Kosińska 2023). Thus it is reasonable to follow this path to model stress barriers of arbitrary forms.

Return to the scheme of Fig. 1 for a fracture tip starting penetration into a barrier. The penetration is hampered by the stress-contrast $\Delta\sigma$, which, for a high barrier, is much greater than the tensile strength of rock. In terms of the linear fracture mechanics (e.g., Rice 1968), this means that absolute value of the (negative) SIF $K_{I\Delta\sigma}$ may dominate in the combined resistance $K_{IR} = (-K_{I\Delta\sigma}) + K_{IC} + K_{IA}$ as compared with the inputs of the fracture toughness $K_{IC}$ and the apparent viscous SIF $K_{IA}$.

Denote $K_I$ the SIF, generated by the driving force which is the net-pressure $p_{net}$. The fracture propagation into a barrier is possible when $K_I$ is equal (for stable fracture growth) or exceeds (for unstable, jump-like growth) the combined resistance $K_{IR}$. The expected fracture growth is stable due to the need to adjust the distance $\Delta z$ of the penetration to a current driving SIF $K_I$. Thus the fracture condition, defining penetration, is

$$(-K_{I\Delta\sigma}) + K_{IA} + K_{IC} = K_I \qquad (16)$$

This is the *key equation* for tracing penetration into a barrier. It is formulated in terms of asymptotic characteristics of fields near a fracture tip. Thus using (16) will be called the *asymptotic approach*.

***Asymptotic solution for penetrable high barriers.*** Specify the terms entering (16) for a general case of stress barrier.

*Driving SIF $K_I$.* According to the general theory (e.g., Muskhelishvili 1953), equation (16) implies that, when neglecting the toughness $K_{IC}$ and apparent viscous $K_{IA}$ SIFs, the normal stress ahead of a crack tip is finite, while the opening goes to zero as $w = O(r^{3/2})$, where $r$ is the distance from the tip. However, this asymptotics holds merely at a very small distance from a tip. Beyond it, the next (square root) term of the asymptotics becomes leading. This asymptotics, conventional in the linear fracture mechanics (e.g., Rice 1968; Murakami 1990), holds at the distance $r$ much greater than an initial penetration $\Delta z$. The distance may be large enough to include a number of grid cells even for quite rough discretization normally used for HF modeling. Thus, for a small penetration, the square-root asymptotics connects the opening with the driving SIF by the classical dependence commonly used in papers on HF modeling:

$$w(r) = \sqrt{\frac{32}{\pi}} \frac{K_I}{E'} \sqrt{r} \qquad (17)$$

By (17), the driving SIF $K_I$ may be evaluated via a fracture opening $w(d)$ at a distance $d$ from the barrier as $K_I = \sqrt{\frac{\pi}{32}} E' \frac{w(d)}{\sqrt{d}}$. This equation is used in computational and experimental fracture mechanics to find the SIF and/or its critical value $K_{IC}$. For our theme, it is reasonable to use its integral form, adjusted to accounting for the mass conservation law which is exactly met by the most schemes of HF modeling. To this end, the driving SIF is expressed in terms of the volume $V_{\Delta s}$ of incompressible fluid, filling a near-barrier rectangular zone. The rectangle has the longer side $r = d$ orthogonal to the barrier, and the side $\Delta s$ along it. Thus $V_{\Delta s} = w_{av}\Delta s d$, where $w_{av}$ is the average opening over the rectangle. A set of such rectangles may be considered as a grid along the barrier. Actually, this means applying a P3D model to a near barrier strip. Integration (17) over the rectangle and solving the result in $K_I$ gives

$$K_I = \frac{3}{2}\sqrt{\frac{\pi}{32}} E' \frac{w_{av}}{\sqrt{d}} \qquad (18)$$

With growing distance $d$, the accuracy of (18) is notably better than that of the starting equation. This becomes evident when employing the most unfavorable (maximal) values of the distance $d$ for a straight or penny-shaped crack under uniform pressure $p_{net}$. By using analytical solutions for SIFs and openings in these cases, it may be seen that for the most unfavorable distance $d = x_*$, the error of (18) is about two-fold less than the error of the starting equation (17). Thus, in further discussion, we shall use $K_I$ defined by (18).

*SIF of stress-contrast $K_{I\Delta\sigma}$.* For a small penetration $\Delta z$, the asymptotic picture corresponds to a plane-strain problem for a semi-infinite crack. Then the asymptotic expression for $K_{I\Delta\sigma}$ is (e.g., Murakami 1990):

$$K_{I\Delta\sigma} = -\sqrt{\frac{8}{\pi}} \Delta\sigma \sqrt{\Delta z} \qquad (19)$$

It can be seen, that with growing stress contrast $\Delta\sigma$, the term $K_{I\Delta\sigma}$ in (16) becomes dominating.

*Tensile critical SIF.* The tensile critical SIF $K_{IC}$ is an assigned characteristic of a material. It is scale dependent, and in field conditions, its value is quite uncertain. For rocks, it normally does not exceed $1\ MPa\sqrt{m}$. For a high barrier, its input into the condition (16) is much less than that of the SIF of stress contrast (19). For instance, in the example of high barrier, given by Chen et al. (2020), the stress-contrast was high when $\Delta\sigma = 4\ MPa$; the grid size was $\Delta x = 2.5\ m$. Hence, when modeling the penetration to the accuracy of the mesh size ($\Delta z = \Delta x = 2.5\ m$), the resisting SIF (19) was $-K_{I\Delta\sigma} = 10.1\ MPa\sqrt{m}$. Therefore, in the problem considered, the input of material resistance $K_{IC}$ may be neglected.

*SIF of viscous resistance $K_{IA}$.* As mentioned, the apparent viscous SIF $K_{IA}$ is defined by the correspondence principle (Linkov and Markov 2020). The latter connects the SIF $K_{IA}$ with the propagation speed $v_*$. Since $v_* = \frac{d\Delta z}{dt}$, using the correspondence principle to find $K_{IA}$ makes the condition (16) an ordinary differential equation (ODE), defining the penetration $\Delta z(t)$ and, consequently $v_*(t)$, as functions of time. At a high barrier, the penetration speed $v_*$ drastically drops, which results in the drop of viscous resistance. By the correspondence principle, for zero penetration speed $v_*$, the apparent SIF $K_{IA}$ is zero, as well. Hence with growing stress-contrast, when penetration speed goes to zero, the viscous term in (16) becomes negligible.

Summarizing, for a high barrier, both the tensile critical SIF $K_{IC}$ and the apparent viscous SIF $K_{IA}$ may be omitted in the penetration equation (16). The equation becomes

$$-K_{I\Delta\sigma} = K_I \qquad (20)$$

It is algebraic. Using (19) in (20) gives the penetration $\Delta z(t)$ as a function of the driving SIF $K_I(t)$:

$$\Delta z(t) = \frac{\pi}{8}\left(\frac{K_I(t)}{\Delta\sigma}\right)^2 \qquad (21)$$

Substitution (18) into (21) gives the general solution for the penetration $\Delta z(t)$ into a high barrier as a function of $w_{av}(t)$:

$$\Delta z(t) = \left(\frac{3\pi}{32} \frac{E'}{\Delta\sigma} \frac{w_{av}(t)}{\sqrt{d}}\right)^2 \qquad (22)$$

Recall that the opening $w_{av}$ used in (22) is averaged over the near barrier part of the fracture surface, where the square-root asymptotic is acceptable. When this part is large, the opening averaged over it, may have the order of the average opening entering equation (9) for the Courant time $cou$. Then large values of $w_{av}$, calculated with spatial discretization in the near-barrier zone may also serve for conclusions on the computational efficiency and stability of a conventional scheme. Thus in general, for high barriers ($R_0 >$ 3.5) the asymptotic approach reduces to using the simple dependence (21) (when calculating the SIF), or (22) (when evaluating the average opening).

**2.4.2 Bench-mark problems**

*Formulation of bench-mark problems for HF penetration into stress barrier.* The approximate solution (22) is derived by using for the driving and resisting SIFs their asymptotic forms (18) and (19). In cases of plane-strain and axisymmetric problems, its accuracy may be checked by comparing with the results obtained with using exact equations for these SIFs. For generality, we shall not neglect the inputs of the toughness $K_{IC}$ and viscosity $K_{IA}$ terms in the key equation (16). Inclusion them into the analysis will serve us to clarify details of the rapid (nearly instant) speed drop immediately after reaching a high barrier, and to thoroughly compare the asymptotic solution with the solution obtained by spatial discretization.

Consider the symmetric scheme (Fig. 2a) for a straight fracture of length $2x_*(t)$ with tips penetrating into barriers with stress contrast $\Delta\sigma$. The boundaries of the barrier are at the distance $x_B$ from the origin, where a pointed source with the pumping rate $Q_0$ is located. The penetration $\Delta z(t)$ into the barrier is $\Delta z(t) = x_*(t) - x_B$. By symmetry, it is sufficient to consider the right part of the picture, for which the fluid influx is $Q_0/2$. The exact equations for SIFs entering (16) are given explicitly in the paper (Linkov and Markov 2020). There is no need to reproduce them because merely the resulting ODE is of interest for our theme.

Omitting technical details, the resulting ODE is

$$\frac{dy}{dt'} = \gamma_x \left[ \left(\frac{1}{1+y}\right)^{\alpha_x} \left(1 + t' - \frac{2\sqrt{2}}{\pi} R_0 \sqrt{y + y^2/2}\right) - \left(\frac{1}{1+y}\right)^{\beta} \frac{K_{IC}}{K_{IB}} \right]^{\omega} \quad (23)$$

where $y = \frac{\Delta z}{x_B}$ is the normalized penetration; $t' = \frac{t-t_B}{t_B}$ is the normalized time, counted from the time of reaching the barrier; $R_0$ is the *starting intensity* of the barrier at this moment. The exponents entering (23) are $\gamma_x = \frac{n+1}{n+2}$, $\alpha_x = \frac{n+4}{n+2}$, $\beta = \frac{1-0.5n}{n+2}$, $\omega = \frac{n+2}{n}$; they depend merely on the fluid behavior index $n$. The ODE (23) is to be solved under the initial condition of zero penetration at $t' = 0$:

$$y(0) = 0 \quad (24)$$

The normalized variables in (23), (24) are defined as the ratios

$$y(t) = \frac{\Delta z(t)}{x_B} = \frac{x_*(t) - x_B}{x_B}, \quad t' = \frac{\Delta t}{t_B} = \frac{t - t_B}{t_B} \quad (25)$$

Recall that $x_B$ is the half-distance between the barriers, $t_B$ is the time of reaching the barriers. Besides, the derivation employs the average opening $w_B$, net-pressure $p_B$ and stress intensity factor (SIF) $K_{IB} = p_B\sqrt{\pi x_B}$ at the instant $t_B$. With $x_B$ being assigned, other normalizing quantities are known from the self-similar solution by Adachi and Detournay (2002) to the classical KGD problem. They are

$$t_B = t_n \left(\frac{x_B^2}{\xi_{*n}^2 Q_0 t_n}\right)^{1/(2\gamma_x)}, \quad w_B = \frac{Q_0 t_B}{2x_B}, \quad p_B = \frac{E'}{\pi} \frac{w_B}{x_B}, \quad K_{IB} = p_B\sqrt{\pi x_B}, \quad v_B = \gamma_x \frac{x_B}{t_B} \quad (26)$$

where $\xi_{*n}$ is the self-similar fracture half-length, depending merely on the fluid behavior index $n$; it is tabulated in the cited paper; its values slowly increase with decreasing $n$ from $\xi_{*n} = 0.615$ for a Newtonian fluid ($n = 1$) to $\xi_{*n} = 0.654$ for a perfectly plastic fluid ($n = 0$).

The normalized start velocity is $\frac{dy}{dt'}\Big|_{t'=0} = \gamma_x$ to the accuracy of the ratio $\frac{K_{IC}}{K_{IB}}$, which is neglected at $t \leq t_B$ *before* reaching the barrier in the viscosity dominated regime. The ratio may become of essence only well *after* reaching the barrier (at $t \gg t_B$), when the regime turns to the toughness dominated due to significant drop of the propagation speed. In terms of physical quantities, the equality $\frac{dy}{dt'}\Big|_{t'=0} = \gamma_x$ expresses that, as should be, the propagation speed $v_* = \frac{d\Delta z}{dt}$ at the moment of reaching the barrier equals to the speed $v_B = \gamma_x \frac{x_B}{t_B}$, defined by the self-similar solution.

Thus the problem is reduced to the Cauchy problem (23), (24). It contains merely three external parameters; these are the starting barrier intensity $R_0 = \frac{\Delta\sigma}{p_B}$, the fluid behavior index $n$, and the ratio $\frac{K_{IC}}{K_{IB}}$. The solution $y(t')$ to the Cauchy problem (23), (24) is promptly found numerically by using a standard subroutine. When having $y(t')$, the normalized average opening $W$ and normalized net-pressure $P$ become known, as well. Omitting again technical details, they are

$$W = T\varsigma, \quad P = T\varsigma^2 + \frac{2}{\pi} R_0(\arccos\varsigma - \varsigma\sqrt{1-\varsigma^2}) \tag{27}$$

where

$$W = \frac{w_{av}}{w_B}, \quad P = \frac{p_{net}}{p_B}, \quad T = \frac{t}{t_B} = 1 + t', \quad \varsigma = \frac{1}{X}, \quad X = \frac{x_*}{x_B} = 1 + y \tag{28}$$

Note that the first of (28) actually expresses the mass conservation law written in the normalized variables. It shows that for small penetration ($y \ll 1$), the average opening is proportional to time.

*Remark on axisymmetric problem.* Similar to the modified KGD scheme (Fig. 2a), the modified axisymmetric scheme (Fig. 2b) is one-dimensional with $x$, $x_*(t)$ and $x_B$ replaced now with polar radii $r$, $r_*(t)$ and $r_B$. As a result, we arrive at the same Cauchy problem (23), (24) with changes in constants entering the normalizing quantities (26) and in numerical coefficients. Specifically, they are changed to

$$t_B = t_n \left(\frac{r_B^3}{\xi_{*n}^3 Q_0 t_n}\right)^{1/(3\gamma_r)}, w_B = \frac{Q_0 t_B}{\pi r_B^2}, p_B = \frac{E'}{\pi}\frac{w_B}{r_B}, K_{IB} = \frac{2}{\sqrt{\pi}} p_B \sqrt{r_B}, v_B = \gamma_r \frac{r_B}{t_B},$$

$$\gamma_r = \frac{2}{3}\frac{n+1}{n+2}, \alpha_r = 2\frac{n+3}{n+2}, \beta = \frac{1-0.5n}{n+2}, \omega = \frac{n+2}{n}$$

where $\xi_{*n}$ is tabulated in the paper (Linkov 2016a); its values slowly increase with decreasing $n$ from $\xi_{*n} = 0.698$ for a Newtonian fluid ($n = 1$), to $\xi_{*n} = 0.733$ for a perfectly plastic fluid ($n = 0$).

The variables $y = \frac{\Delta z}{r_B}$, $t' = \frac{t-t_B}{t_B}$ are defined as in (23) by (25). In view of the entire analogy with the plain-strain case, it is sufficient to consider the latter. The normalized average opening $W = \frac{w_{av}}{w_B}$, and the normalized net-pressure $P = \frac{p_{net}}{p_B}$ for the axisymmetric problem are

$$W = T\varsigma^2, \quad P = T\varsigma^3 + R_0(1-\varsigma^2)^{3/2}$$

with $T$ and $\varsigma$ defined in (28).

*Numerical results and their interpretation.* Both for plane-strain and axisymmetric problems, the Cauchy problem (23), (24) contains merely three external parameters; these are the fluid behavior index $n$, the ratio $\frac{K_{IC}}{K_{IB}}$ and the starting intensity of the stress contrast $R_0 = \frac{\Delta\sigma}{p_B}$. It is promptly solved by a standard solver when the behavior index $n$ is non-zero. If $n = 0$, the problem may be reformulated accounting for degeneration of the Poiseuille-type equation in this case. For certainty, we focus on the plane-strain problem and commonly considered case of a Newtonian fluid ($n = 1$). Then $\gamma_x = 2/3$, $\alpha = 5/3$, $\beta = 1/6$, $\omega = 3$, $t_n = \mu'/E'$, $\xi_{*n} = 0.615$. When neglecting the toughness $K_{IC}$, the ODE (23) contains the only parameter, the starting intensity of stress contrast $R_0$. We present its bench-mark solutions for various $R_0$.

Integration the ODE (23) under the initial condition (24) was performed by using the standard Fortran subroutine IVPRK. It solves a Cauchy problem by the Runge-Kutta-Verner fifth-order method to an assigned tolerance. The tolerance was set as $TOL = 5 \cdot 10^{-4}$. The calculations covered six orders time interval. The corresponding normalized time $t' = \frac{t-t_B}{t_B}$ changed from $0.5 \cdot 10^{-3}$ to $0.5 \cdot 10^3$. To preserve the accuracy on the tolerance level, the calculations were performed in a number of stages with using an output of a finer time scale, to assign the initial condition for the next rougher scale. When having the normalized penetration $y(t')$, the normalized average opening $W$ and net-pressure $P$ are found by means of (27). (In the case of the axisymmetric problem, they are given at the remark above).

The calculated dependencies of the normalized penetration $y$, penetration speed $V = dy/dt'$, opening $W$ and net-pressure $P$ on the normalized time $t'$, are presented in Fig. 3. The normalized time, normalized penetration and speed, which change some orders, are given in logarithmic scales.

a)

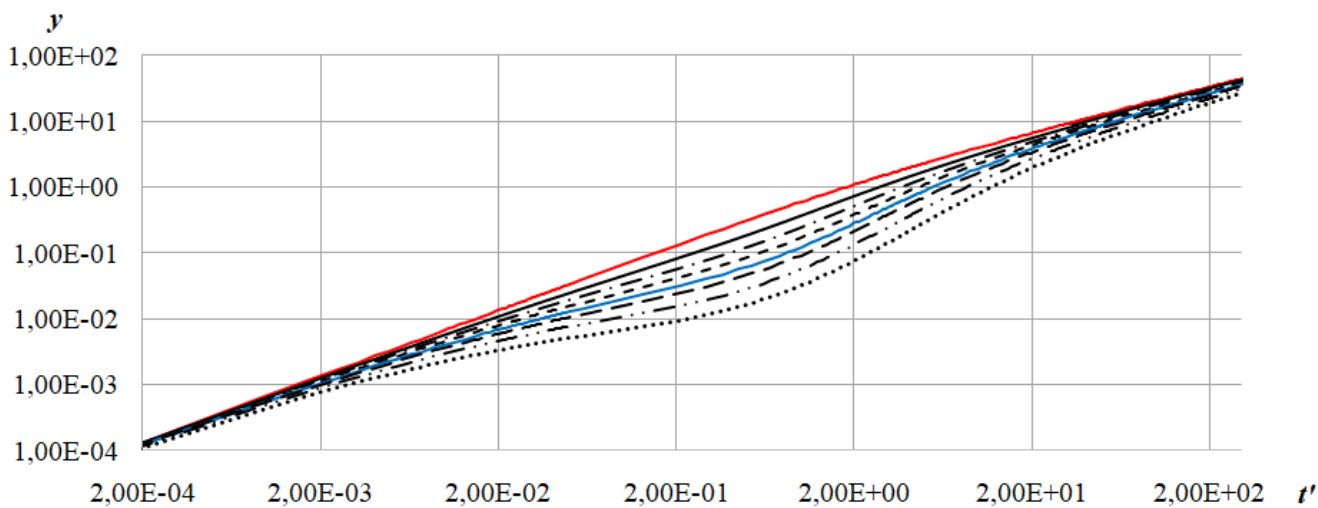

b)

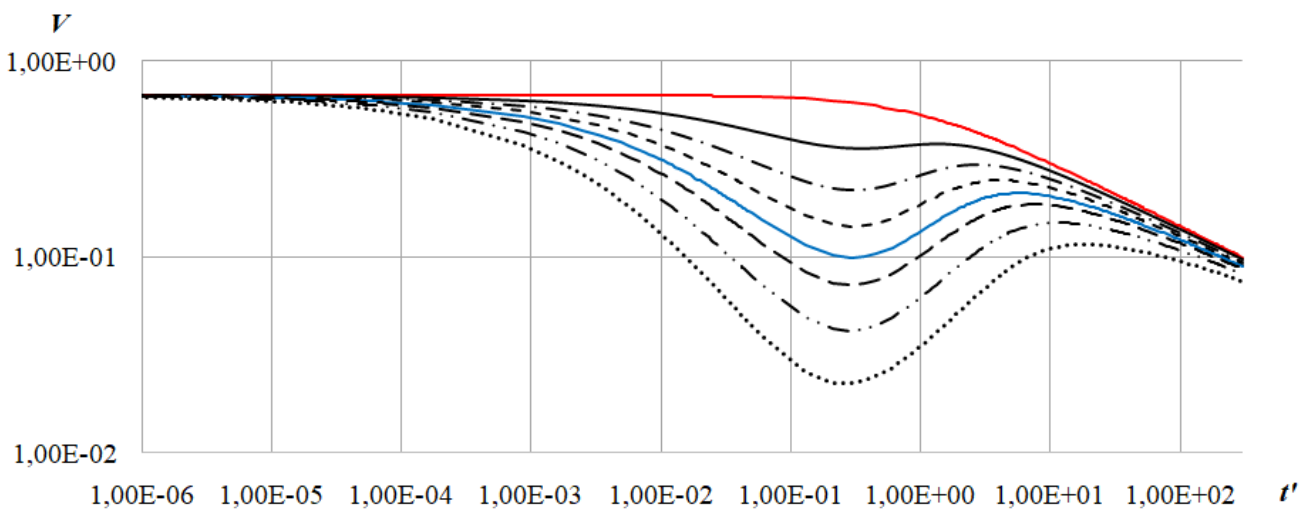

c)

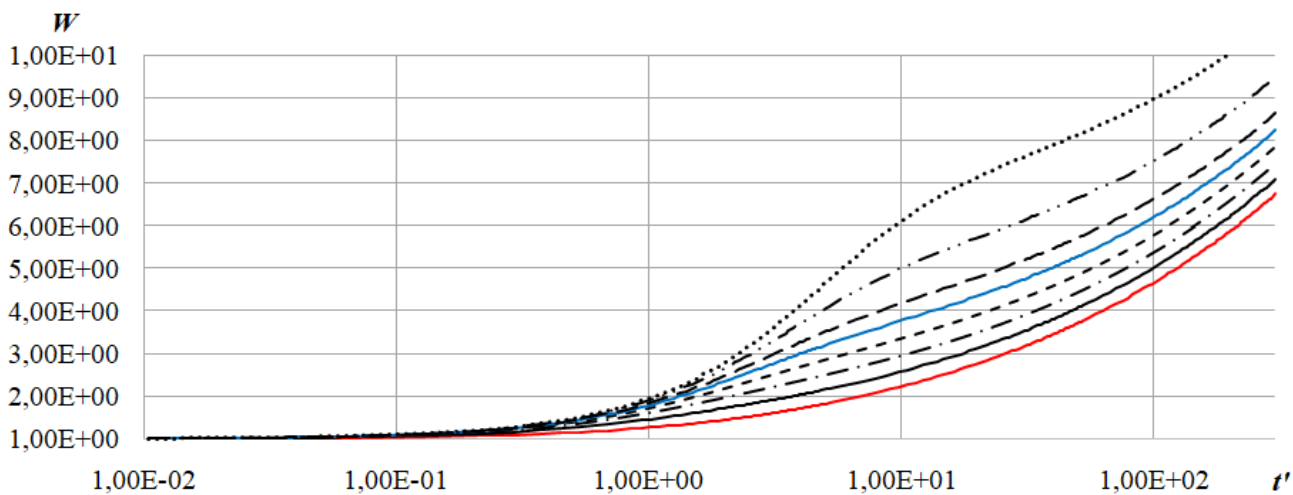

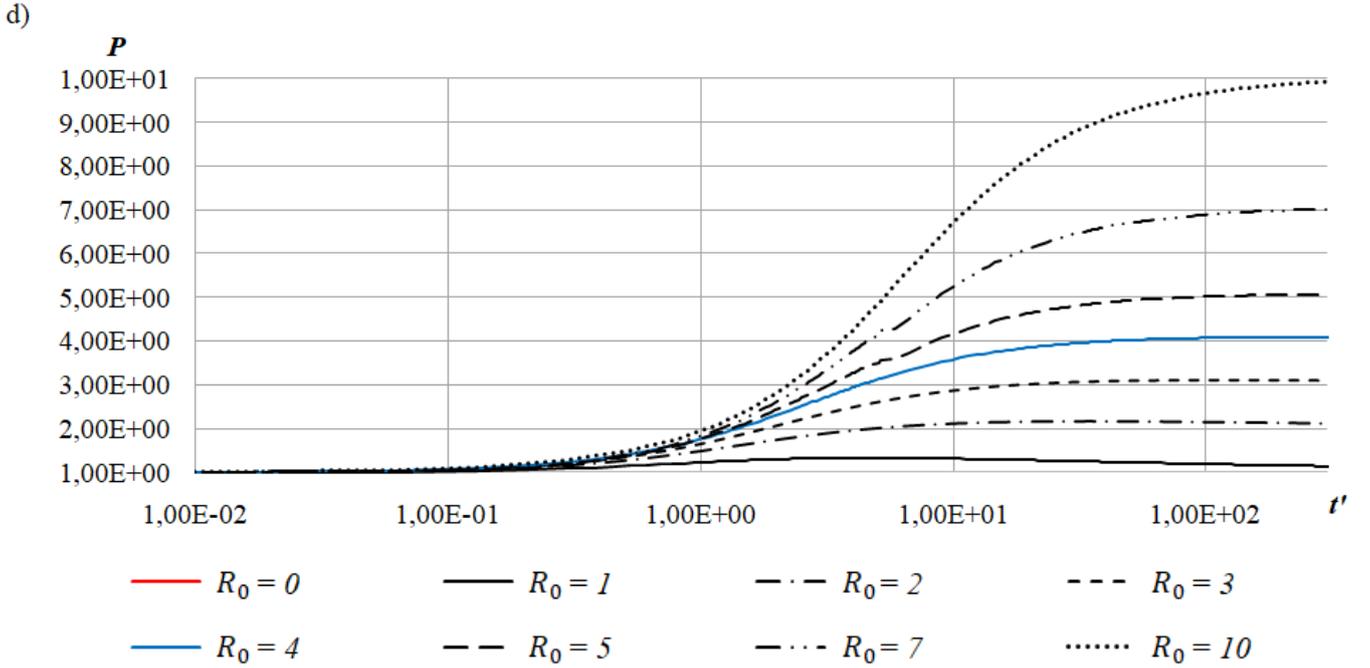

**Fig. 3** Dependencies of (a) the normalized penetration $y$, (b) speed $V$, (c) opening $W$ and (d) pressure $P$ on the normalized time $t'$ for various values of the barrier intensity

The graphs quantitatively illustrate the expected physical features of the penetration into a barrier. The curves for the speed (Fig. 3b) are especially instructive. They show that there are three distinct stages of the penetration. Firstly, the speed decreases due to hampering influence of a barrier. On the second stage, the speed increases due to the growth of the average net-pressure (Fig. 3d), caused by the barrier. On the third stage, the speed again decreases. The decrease starts when the net-pressure increases to the level of the stress contrast, so that the *current* barrier intensity $R(t) = \Delta\sigma/p_{net}(t)$ becomes about 1.4. The decrease of the speed occurs due to turning to the *asymptotic regime*, corresponding to the propagation under dominating influence of the new *nearly uniform* in-situ stress. Recall that, while the starting in-situ stress between the barriers is $\sigma_0$, the in-situ stress behind the barriers is $\sigma_0 + \Delta\sigma$. Thus, with growing length of the fracture, the net-pressure is to be defined with respect to the latter value. In limit, the propagation corresponds to that in a medium with uniform in-situ stresses $\sigma_0 + \Delta\sigma$. Consequently the solution tends to the dependence, corresponding to the solution for the case $R_0 = 0$, studied by Adachi and Detournay (2002). In particular, for a Newtonian fluid, this solution yields that the speed decreases as $t^{-1/3}$ with growing time. Such limiting behavior clearly appears in Fig. 3a-c, which show that the solutions for $R_0 = 0$ envelope all the curves for various starting intensities $R_0$, when the time grows.

*Nolte-Smith parameter and arrest time.* Recall that the change of the pressure in time is of special value for conclusions on the fracture propagation (e.g., Nolte and Smith 1981; Nolte 1989; Linkov, Rybarska-Rusinek and Rejwer-Kosińska 2023). Thus, it is reasonable to study the dependence $P(t')$ in more detail. From the second of (27), it may be inferred that in the extreme cases of zero ($R_0 = 0$) and infinite ($R_0 = \infty$) intensities of the barrier, the dependences of the pressure on the time become monomial $P = T^{b_P}$ with the Nolte-Smith (1981) slope parameter $b_P = \frac{d\log P}{d\log T}$ equal to $-1/3$ for $R_0 = 0$, and $1.0$ for $R_0 = \infty$. Furthermore, writing the second of the dependencies (27) as

$$\frac{P}{T} = \frac{P}{1+t'} = \varsigma^2 + \frac{1}{\sqrt{2a}}(\arccos\varsigma - \varsigma\sqrt{1-\varsigma^2}) \qquad (29)$$

with

$$a = \left(\frac{\pi}{2\sqrt{2}}\frac{T}{R_0}\right)^2 = \left(\frac{\pi}{2\sqrt{2}}\frac{1+t'}{R_0}\right)^2 \qquad (30)$$

and expanding the solution in series in $a$, yields that to the accuracy of $O(a^3)$, the ratio $\frac{P}{T}$ is approximated by the second order polynomial

$$\frac{P}{T} = \frac{P}{1+t'} \approx 1 - \frac{2}{3}a + \frac{4}{5}a^2 \quad (31)$$

The derivation has assumed that the normalized penetration $y$ or/and the normalized speed $V$ are sufficiently small as compared with the unity.

For the Nolte-Smith slope parameter $b_P$, the approximation (31) yields

$$b_P = \frac{d\log P}{d\log T} = 1 - 4a\frac{5/12 - a}{5/4 - 5/6a + a^2} \quad (32)$$

The maximal value of the slope parameter corresponds to an impenetrable barrier and equals to 1.0. This implies that the nominator on the right hand side of (32) cannot be negative. Hence, the approximate equation (31) and its corollary (32) are applicable only in the interval $0 \le a \le 5/12$. In this range, the slope parameter changes from its maximal value $b_P = 1.0$ when $a = 0$, through the minimum $b_P = 0.845$ reached at $a = 3.0 - \sqrt{7.75} = 0.2161$, back to the maximum $b_P = 1.0$ for $a = 5/12$. Therefore, in the whole range considered, the slope parameter is close to its maximal value $b_P = 1.0$, corresponding to impenetrable barrier. By the definition (30), the corresponding interval of the normalized time $t'$ is $0 \le t' \le t'_A$, where

$$t'_A = \begin{cases} 0 & \text{when } 0 < R_0 \le 1.724 \\ 0.581 R_0 - 1.0 & \text{when } R_0 > 1.724 \end{cases} \quad (33)$$

The time $t'_A$ is the maximal normalized time during which the barrier may be considered practically impenetrable ($b_P \approx 1.0$). This time may be associated with the *arrest time*, caused by the barrier for the fracture propagation. The value $R_0 = 1.724$ is found from (30) for the maximal acceptable value of $a$ ($a = 5/12$) and the minimal possible value of the time ($t' = 0$). From (33) it follows that for $R_0 > 1.724$, the arrest time $t'_A$ linearly grows with the increase of the starting barrier intensity $R_0$.

With these prerequisites, we may compare the exact solution for the pressure history $P(t')$, defined by (29), with the approximation (31). The comparison is given in Fig. 4. The arrest times $t'_A$, defined by equation (33), are shown by crosses for each of the starting barrier intensities $R_0$.

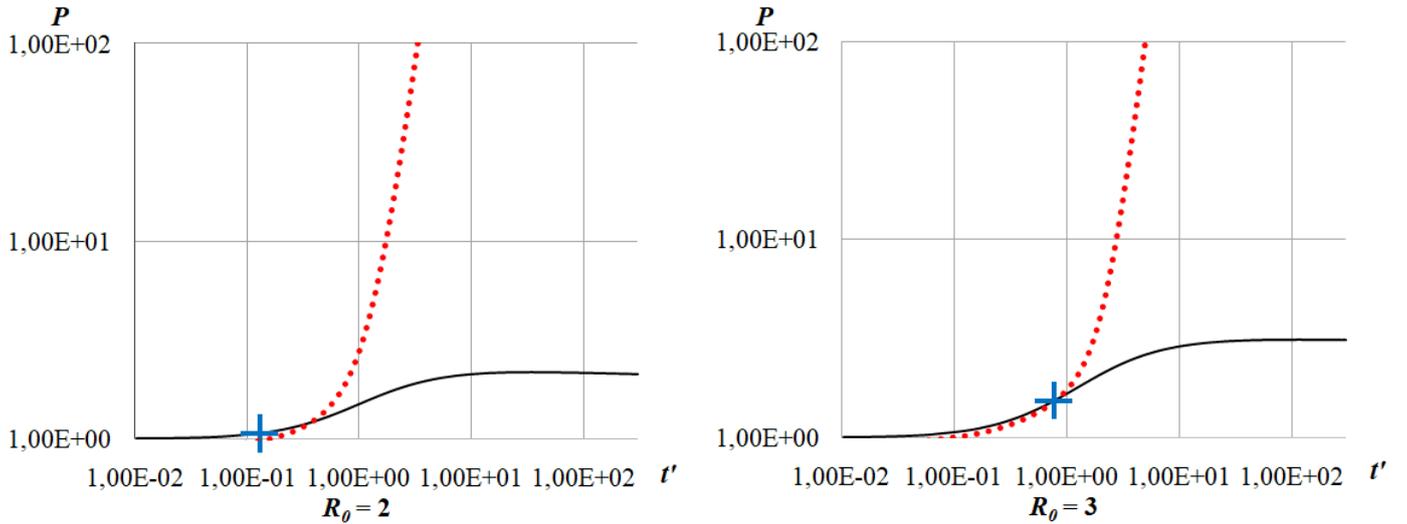

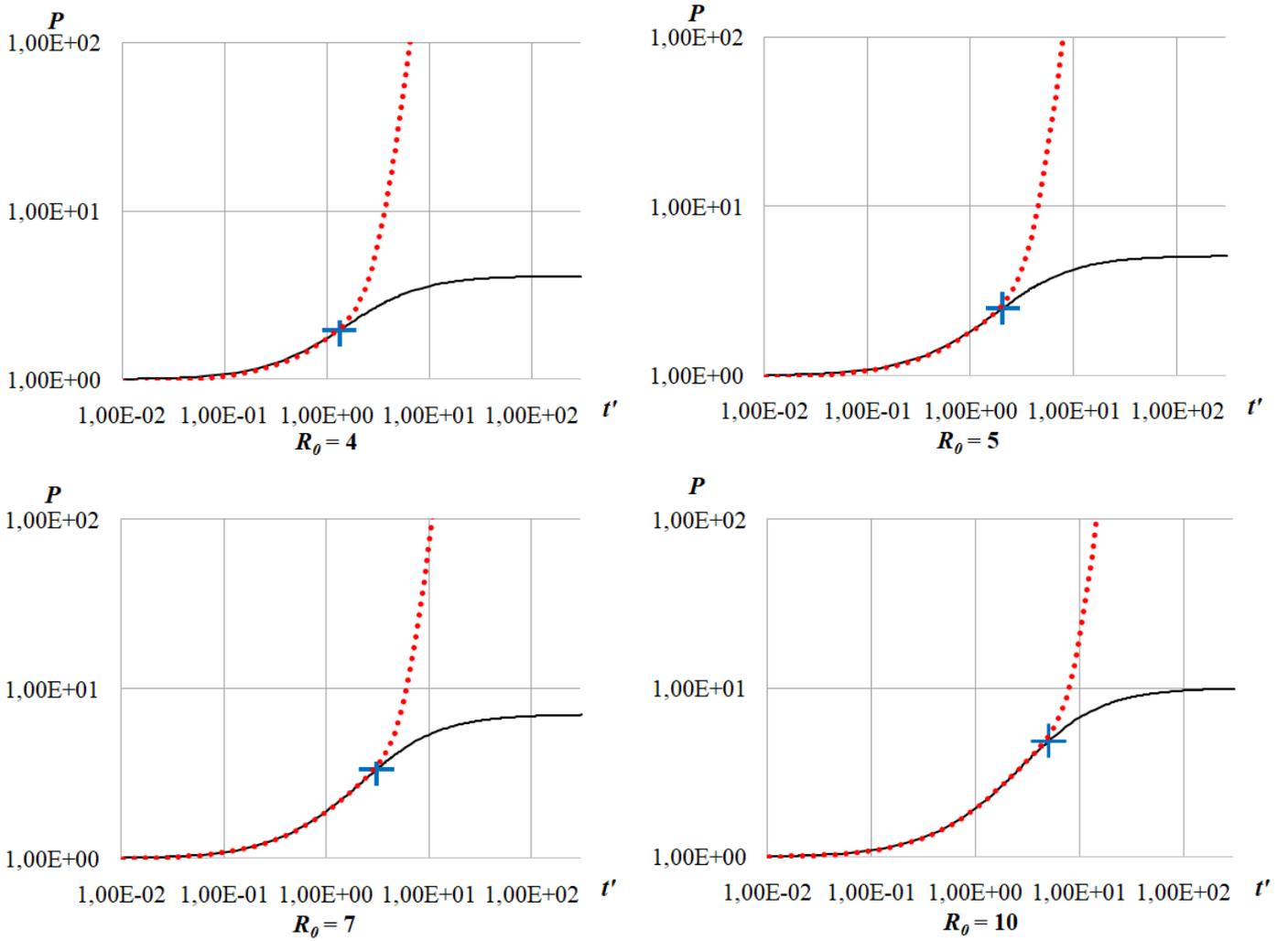

**Fig. 4** Comparison of the calculated normalized pressure histories (solid lines) with their approximation (31) (dotted lines) for various values of the barrier intensity ($R_0 = 2; 3; 4; 5; 7; 10$)

It can be seen that the approximation (31) is quite accurate in the entire range of the normalized time $0 \leq t' \leq t'_A$ from reaching the barrier till the arrest time. After that, the approximation drastically deteriorates. As mentioned, the arrest time $t'_A = 0.58 R_0 - 1.0$ linearly grows with the barrier intensity from zero, when the barrier is weak ($R_0 < 1.72$), so that it does not influence the fracture propagation, to values exceeding 1.0 for a sufficiently strong barrier ($R_0 > 3.5$).

As appears from Fig. 3a, the normalized penetration $y$ during the arrest time is quite small, being less than 0.1 for $R_0 = 2$ and decreasing to less than 0.01 for $R_0 = 7$. Consequently, $\varsigma = 1/(1+y) \approx 1$, and by the mass conservation law, the normalized opening, defined in (27), grows during the arrest time proportionally to the normalized time: $W = T = 1 + t'$. Since $w = W w_B$ and $w = T t_B$, this implies that for strong barriers, the cubed opening rapidly grows in time; in limit it goes to infinity with the time growth. Then, as clear from equations (9), (12), (14), (15), the stability and efficiency of any scheme using spatial discretization would deteriorate. Below, we shall illustrate this by considering the results of the paper (Gladkov and Linkov 2018), obtained for the plane-strain problem solved with spatial discretization.

Within the arrest interval $0 \leq t' \leq t'_A$, the Nolte-Smith slope parameter is close to its maximal value $b_P = 1.0$, corresponding to impenetrable barrier. When the normalized time exceeds the arrest time, the slope parameter $b_P$ rapidly decreases, what manifests overcoming the barrier. These analytical and numerical results show the significance of the slope parameter for fracture propagation through barriers. They once again confirm the prophetic assertion by Nolte (1989, p. 304): "…the log-log plot of net pressure vs time is a basis for interpreting pressures during fracturing."

## 3 PART II. Accuracy and range of applicability of asymptotic approach

### 3.1 Limiting cases of very high and nearly zero stress contrasts

The Cauchy problem (23), (24) has simple analytical solutions in the cases of very high ($R_0 \to \infty$) and very small ($R_0 \to 0$) stress barriers. In the first case, the starting equation (16) reduces to (20). Then substitution into (20) the analytical expressions for driving and stress-contrast SIFs, known for a straight crack from the Muskhelishvili (1975) solution and used to derive (23), yields that when $t > t_B$ the relative penetration is

$$\frac{\Delta z}{x_*(t)} = 1 - \cos\left(\frac{\pi}{2}\frac{p_{net}(t)}{\Delta\sigma}\right) \approx \frac{\pi^2}{8}\left(\frac{p_{net}(t)}{\Delta\sigma}\right)^2 \tag{34}$$

Equation (34) actually reproduces the classical formula by Dugdale (1960) for the size of a plastic zone at the crack tip. Now $p_{net}$ replaces the tensile stress and $\Delta\sigma$ replaces the yield stress. The Dugdale's formula is one of the most famous achievements in fracture mechanics due to accurate prediction of the size of the plastic zone measured in experiments. Equation (34), obtained in two quite different instances, reflects the analogy between their physical pictures. The fact that the Dugdale's formula is well-established suggests that in cases of high stress barriers ($R_0 \gg 1$), the mathematical model developed above is reliable, as well.

In the opposite extreme case of small, in limit zero barrier intensity ($\Delta\sigma = 0; R_0 = 0$), ODE (23) becomes

$$\frac{dy}{dt'} = \gamma_x \left[\left(\frac{1}{1+y}\right)^{\alpha_x}(1+t')\right]^\omega \tag{35}$$

In the axisymmetric case, the exponents $\gamma_x$ and $\alpha_x$ are replaced to $\gamma_r$ and $\alpha_r$, respectively. It is easy to check by direct substitution that the exact solution to (35) under the Cauchy condition (24) is $y = (1+t')^{\gamma_x} - 1$. In fact, for the plane-strain problem, it is the classical self-similar solution by Adachi and Detournay (2002), written in the normalized variables $y$ and $t'$. Similarly, for the axisymmetric problem, the exact solution $y = (1+t')^{\gamma_r} - 1$ is the self-similar solution given in (Linkov 2016a). Thus, for $R_0 = 0$, the solution to the Cauchy problem (23), (24) is the known self-similar solution.

Summarizing, the solution to the Cauchy problem (23), (24) provides physically sustainable and accurate results in the limiting cases of very high ($R_0 \gg 1$) and very low ($0 \le R_0 \ll 1$) stress barriers. By continuity, we may expect that the solution is sound in cases intermediate between the extreme.

### 3.2 Comparison with plane-strain solution obtained by spatial discretization

Consider, the example of very high stress barrier studied in the paper (Gladkov and Linkov 2018) by using spatial discretization. All the physical input parameters, except for the intensity of the stress contrast, were the same as in the paper by Dontsov and Peirce (2015): $n = 1$, $\mu' = 1.2\, Pa \cdot s$, $E' = 2.5 \cdot 10^4\, MPa$, $K_{IC} = 0$, $x_B = 25\, m$, $Q_0 = 5 \cdot 10^{-4}\, m^2/s$. For them, the self-similar solution by Adachi and Detournay (2002) yields $t_n = 4.8 \cdot 10^{-11}\, s$, $t_B = 203.904\, s$, $w_B = 2.039 \cdot 10^{-3}\, m$, $p_B = 0.649\, MPa$, $K_{IB} = 5.752\, MPa \cdot \sqrt{s}$, $v_B = 0.0817\, m/s$. The results below refer to the case of the highest stress contrast $\Delta\sigma = 50\, MPa$, studied in (Gladkov and Linkov 2018) with using rather fine grid with 200 cells along a half-length. The corresponding intensity of the barrier is $R_0 = 77.04$.

Present firstly the results obtained by the *asymptotic approach* without computational difficulties. They are given in Figures 5-8 in terms of physical time, penetration, propagation speed and average opening. For the example considered, the stress barrier, while being very high, is *finite*. Thus the propagation speed stays continuous. The calculations show that it drastically drops from the initial value $v_B = 0.0817\, m/s$ to $v_{*min} = 0.600 \cdot 10^{-4}\, m/s$ during the time interval of $12\, s$. Fig. 5 illustrates how fast the drop occurs immediately after reaching the barrier (during the first second).

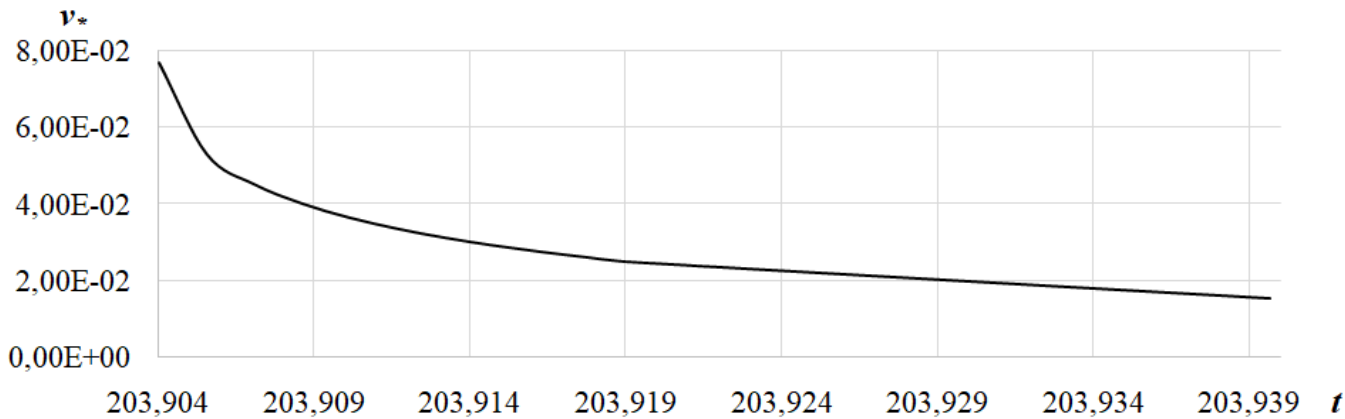

**Fig. 5** Rapid drop of the propagation speed immediately after reaching the barrier at $t_B = 203.904\ s$

The change of the propagation speed on the whole interval of calculations from $t_B = 203.904\ s$ to $t = 11000\ s$ is shown in Fig. 6. It clearly demonstrates that the speed decrease happens practically as a jump. After that, the speed starts to grow. The growth is slow and almost linear on the major part of the further penetration of the tip in the barrier.

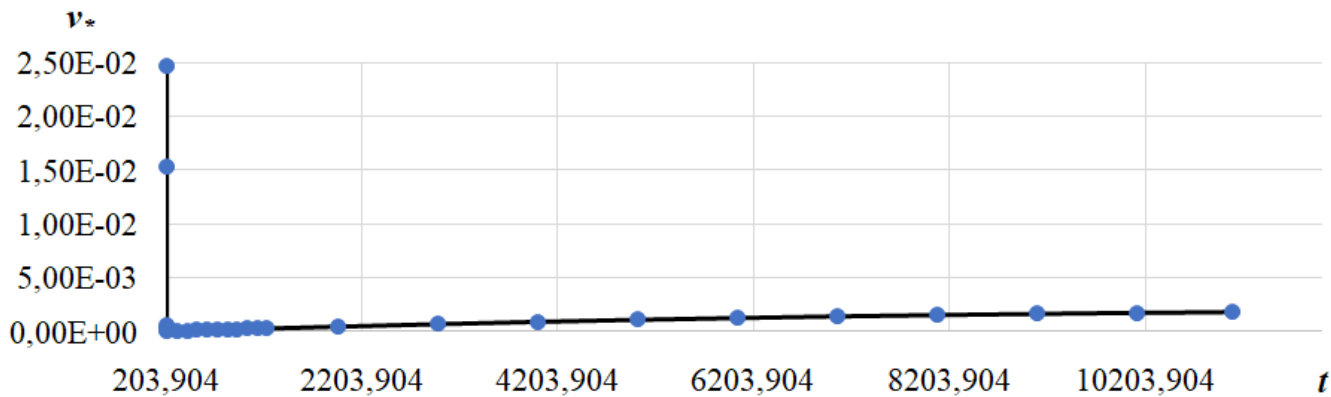

**Fig. 6** Change of the propagation speed on the entire interval of calculations

The tip position is presented in Fig. 7. For the almost linear change of the speed, the penetration $\Delta z$ grows approximately proportional to the squared time elapsed after reaching the barrier.

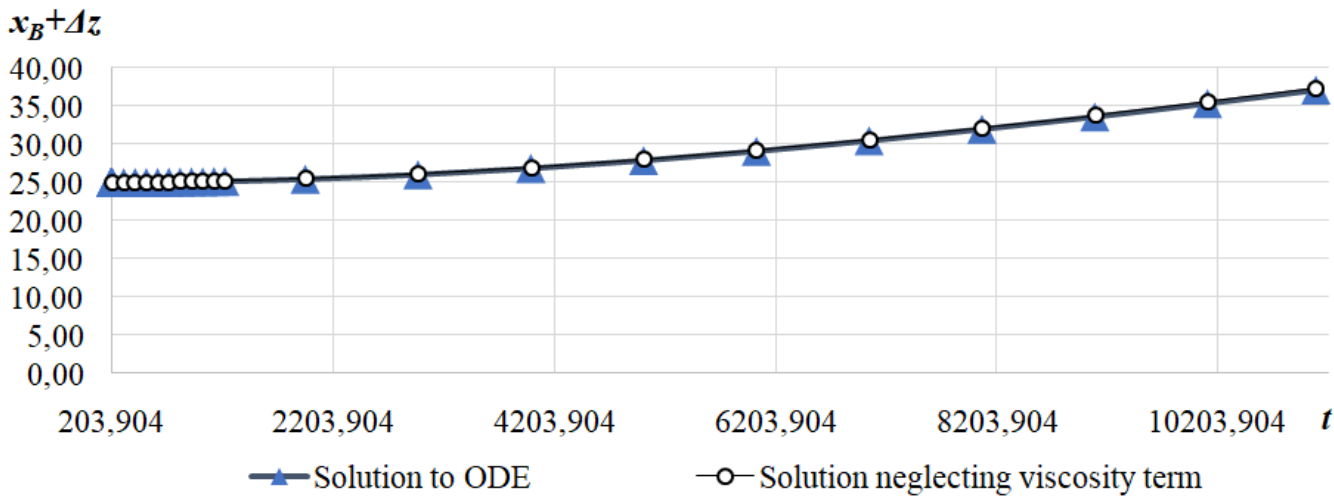

**Fig. 7** Dependence of penetration on time after reaching the barrier

From Fig. 7, it can be seen that 5%-increase of the fracture length occurs only after some $3000\ s$. Its length $2x_*$ is nearly the same as the distance between barriers $2x_B$. Then, by the global mass balance

$w_{av}(t)x_*(t) = \frac{1}{2}Q_0 t$, the growth of the average opening is approximately linear $w_{av}(t) = w_B + \frac{Q_0}{2x_B}(t - t_B)$, The graph of the calculated opening in Fig. 8 evidently agrees with such growth. At the time $t = 3203\ s$, the opening is 15-fold greater than its starting value $w_B = 2.039 \cdot 10^{-3}\ m$. At the arrest time, which is $t_A = 9\ 127\ s$, the opening becomes 32-fold greater. In view of (9), (12) and (14), these estimations indicate there may appear difficulties when using spatial discretization.

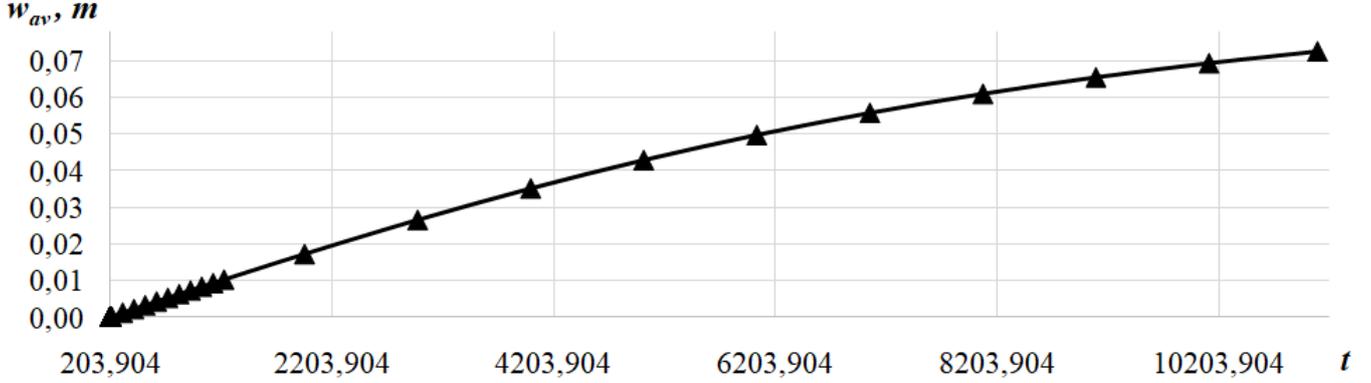

**Fig. 8** Dependence of average opening on time after reaching the barrier

At large time $t = 11000\ s$ (more than three hours), the fracture length $2x_*$ becomes 1.5-fold greater than the distance $2x_B$ between the barriers. The average opening is $w_{av} = 7.47 \cdot 10^{-2}\ m$; it is 37-fold greater than the starting value.

The results presented in figures 6-8 evidently demonstrate that beyond a very short interval immediately after reaching the barrier (Fig. 5), the input of the viscosity term is negligible, and the penetration occurs in the storage dominated regime. Mathematically, this means that on the major part of the penetration into high barrier, the viscosity resistance term $K_{IA}$ may be neglected in the key equation (16). Then ODE (23) becomes an algebraic equation with the solution $y = \sqrt{1 + 2a^2} - 1$, where $a$ is defined by (30). The corresponding dimensional penetration is shown in Fig. 7 by white circles. Actually, in the graphical representation, it is indistinguishable from the exact solution to ODE (23). Similarly, the dependence of the net-pressure on time also practically coincides with the approximation (31) in the whole range from reaching the barrier at $t_B = 203.904\ s$ to the arrest time $t_A = 9\ 127\ s$.

Compare the solution to (23), (24) with the solution (Gladkov and Linkov 2017, 2018), obtained by *spatial discretization*. In the paper cited, the problem was solved for grids having from 20 to 200 nodes at a fracture half-length and for stress contrasts varying from $R_0 = 0$ (classic KGD case), through moderate ($R_0$ from 2 to 3.5) and strong ($R_0 = 10$) barriers, up to extremely high barrier with the mentioned intensity $R_0 = 77.04$. For the finest grid $\Delta x = 1.25 \cdot 10^{-2}\ m$, the Courant time (9) when reaching the barriers was $cou = 1.1 \cdot 10^{-8}\ s$. Since for an explicit scheme of integration, a time step $\Delta t$ cannot exceed $cou$, the step $\Delta t$ used in the paper cited was quite small ($\Delta t < 1.1 \cdot 10^{-8}\ s$). Using so small time step was possible because the problem, being one-dimensional, the order of the matrix resulting spatial discretization was relatively small.

Conventional *explicit* (forward Euler and Runge-Kutta of fourth order) and a number of *implicit* schemes (backward Euler, Petzold–Gear backward differential formula (BDF) and Brayton-Gustavson-Hachtel BDF (Brayton et al. 1972)) were used for time stepping. It has been established, that even for rather high stress contrasts of intensity $R_0 = 10$, all the methods, both explicit and implicit, were stable and provided the same (within the tolerance accepted) numerical results. However, for *extremely high barrier* ($R_0 = 77.04$), studied above by the asymptotic approach, with time growth, the opening becomes 30-fold greater. Thus, the Courant time (9) becomes four orders less. As a consequence, the stability of calculations drastically decreases. As reported in the paper cited, the explicit, as well as the implicit backward Euler and Petzold-Geer BDF methods, failed. In this case, only Brayton-Gustavson-Hachtel BDF method provided stable physically consistent results. As mentioned, it provided results indistinguishable from those of the asymptotic approach. Still, as clear from the specific difficulty discussed in Subsection 2.3, even this extremely stable method would unavoidably fail due to further growth of the opening when considering barriers of higher intensity and greater time intervals.

Of value is also to compare details of the solutions, revealed for the extremely high barrier. Note first that, as mentioned, the error of finding the front position was of the grid size order. Besides, *immediately after* reaching the barrier, the calculated *particle velocity* had high-frequency oscillations with the length of the mesh size $\Delta x$ along the facture (Fig. 4 of the paper (Gladkov and Linkov 2017)). The oscillations rapidly (during $10\ s$) damped practically to zero. Evidently, this was a grid dependent effect. Hence, the solution, obtained by spatial discretization, cannot accurately reproduce the abrupt drop of the propagation speed after reaching the high barrier. *In contrast*, the asymptotic approach, avoiding spatial discretization, provides physically consistent results on the speed drop immediately after reaching the barriers. Fig. 5 above illustrates this. It shows that the speed drops an order during very short interval of $0.04\ s$ after reaching the barriers. In the scale of minutes (Fig. 6), the drop looks instant.

Another noteworthy feature refers to the change of the propagation speed *well after* reaching the barrier. As appears from the paper (Gladkov and Linkov 2017), during $50\ min$ after reaching the barrier, the *propagation speed* has oscillations. Their periods (in average 70 s) look connected with the time interval ($t_B \approx 204\ s$), during which the fracture moved in the viscosity dominated regime, and consequently the propagation was strongly influenced by viscosity. From the first of the definitions (26), it follows that $t_B$ is the only characteristic time of the particular boundary value problem (BVP), in which the half-distance $x_B$ between the barriers is the only assigned linear length. The periods of the oscillations are from one third to one fourth of the time $t_B$. They do not look connected with the mesh size $\Delta x = 0.125\ m$, which was 200-fold less than $x_B = 25\ m$. Thus, there are good reasons to conclude that the oscillations of the propagation speed, revealed by calculations with the fine grid of 200 cells on a half-length, describe a real physical effect caused by the presence of the barrier. The oscillations disappeared for time exceeding 3000 s, when the growth of the net-pressure in time led to notable (an order) decrease of the stress-contrast intensity (from $R_0 = 77.04$ to $R < 5$). Remarkably, for intensities less than 5, the publications on solutions, obtained by spatial discretization (Peirce 2015; Zia and Lecampion 2019; Chen et al. 2020; Linkov, Rybarska-Rusinek and Rejwer-Kosińska 2023), do not report on oscillations of the kind after reaching a stress barrier. Thus, this effect, if it is really caused by the physical reason, is commonly beyond the accuracy of conventional calculations employing spatial discretization. Note that in the limit, when the stress-contrast is zero, the effect should disappear. Indeed, then the solution being self-similar does not contain a characteristic length/time. For similar reason, this effect cannot appear in the asymptotic solution, because the Cauchy problem (23), (24), formulated in non-dimensional quantities, does not contain a characteristic length/time.

### 3.3 Accuracy of asymptotic approach in general case

*Errors of asymptotic approach caused by rough evaluation of driving SIF*. The bench-mark solutions are found by using in (16) the exact formulae for the SIFs $K_I$, $K_{I\Delta\sigma}$, and the average opening $w_{av}$. The solutions may serve to estimate the error of the penetration $\Delta z$ in a general case.

The most unfavorable estimation (17) of the average opening $w_{av}$ corresponds to the maximal value of the distance $r = d$, which is the fracture half-length $x_*$ in the plane-strain and the radius $r_*$ in the axisymmetric case. In the plane-strain case, the worst estimation of the average opening is $w_{av} = \pi x_* \frac{p_{net}}{E'}$. Its substitution into (22) gives $\frac{\Delta z}{x_*} = \left(\frac{3\pi^2}{32}\right)^2 \left(\frac{p_{net}}{\Delta\sigma}\right)^2$. Comparing the factor $\left(\frac{3\pi^2}{32}\right)^2$ with the factor $\frac{\pi^2}{8}$ in equation (21), obtained from the analysis of the exact solution, shows that the approximate numerical factor is 30.6% less. Similar result follows also for the axisymmetric problem.

The given estimation refers to the upper bound of the errors, because we have used the most unfavorable value of the distance $d$. Having in mind that in practical modeling of hydraulic fractures, the driving SIF is found quite inaccurately, the accuracy provided by using (22) appears acceptable.

*Accuracy of asymptotic solution as compared with results of truly 3D modeling.* The conclusion that in the case of closed impenetrable contour, the net-pressure becomes nearly constant suggests that in general, when the stress barrier is high ($R_0 > 3.5$), the net-pressure at a plane-strain rectangular zone, discussed in Subsection 3.3, is nearly constant. Hence, when solving numerically the elasticity equation, the entire rectangle may be taken as a single grid cell. For a barrier with a straight boundary, a set of such rectangles presents a strip, in which the pressure changes in the direction parallel to the boundary. In cases, when there are two high barriers with parallel boundaries, the rectangles with constant pressure elongate up to merging. Then we have merely 1-D grid along barriers, and the scheme becomes that of the P3D model. Detailed truly 3D calculations, performed by (Chen et al. 2020) and reproduced in (Linkov, Rybarska-Rusinek and Rejwer-Kosińska 2023) for a fracture between parallel stress barriers with various intensities, have shown that the net-pressure rapidly becomes constant over a cross section (Fig. 21b of the paper (Chen et al. 2020)). The propagation turns into channelized with the length-to-height ratio about 30 when $R_0 \approx 3.5$. For the greatest of the modelled intensities, $R_0 = 4.66$, the ratio reaches 50. Thus the numerical results of the papers cited may serve to conclude on the accuracy of the approximate equation (22) when the stress contrast is not extremely high.

We shall refer the results summarized in Fig. 19 and 20 of the paper (Chen et al. 2020). Four values of symmetric stress-contrasts were used: $\Delta\sigma = 1.0, 2.0, 3.0, 4.0 \ MPa$. The corresponding intensities were: $R_0 = 1.17, 2.34, 3.50, 4.66$. The grid size was $\Delta x = 2.5 \ m$. Hence, along the half-height $H/2 = 10 \ m$, there were 4 grid cells only. The barriers are reached by the axisymmetric fracture at $t_B = 0.47 t_n \left(\frac{H^3}{Q_0 t_n}\right)^{3/4}$ (Linkov, Rybarska-Rusinek and Rejwer-Kosińska 2023) with $t_n$ defined by (10). In the example considered, $t_B = 2.0 \ s$. At this instant, the calculated net-pressure and consequently the inlet opening were rather uncertain. For the rough mesh used, they became more/less reliably evaluated merely at time at least five-fold greater. Then the net-pressure was nearly uniform with the value $p_B$ of about $1 \ MPa$ in the inlet cross-section; the corresponding inlet opening $w_B$ was about $1 \ mm$ for all the stress barriers studied. Thus, roughly $p_B = 1 \ MPa$, $w_B = 1 \ mm$. At the final time of $t_F = 10 \ min$, the calculated openings were, respectively, $w_F(0) = 3.0, 3.15, 3.20, 3.25 \ mm$ for the stress contrasts listed. The penetration of the fracture into the barrier was $\Delta z_F = 42, 12, 5$ and $3 \ m$.

Use the computed openings $w_F(0)$ in (22) with the most unfavorable $d = H/2$, and take into account that for the plane-strain state, $w_{avF} = (\pi/4)w_F(0)$. This gives $\Delta z_F = 47, 13, 5$ and $3 \ m$ for $\Delta\sigma = 1, 2, 3, 4 \ MPa$, respectively. The agreement with $\Delta z_F = 42, 12, 5$ and $3 \ m$ by Chen et al. (2020) is satisfactory. Thus, despite the estimation (22) is rough, it may serve for practical calculations performed in a conventional way.

## 4 PART III. Practical recommendations

### 4.1 Rough numerical estimations of barrier intensity

Conventional calculations for problems involving stress-contrasts (e.g., Dontsov and Peirce 2015; Gladkov and Linkov 2017; Chen et al. 2020; Linkov and Markov 2020; Linkov, Rybarska-Rusinek and Rejwer-Kosińska 2023) show that fracture growth tends to become channelized when $R_0 > 3.5$. In general, using the barrier intensity (7) may serve to indicate when this occurs. Then, as discussed above, conventional tracing of further propagation may be significantly complicated.

There are two options to use the intensity (7) of a stress barrier, (i) either directly by means of the calculated net-pressure, or (ii) through the average opening $w_{av}$ in the plane-strain zone near the barrier (in particular, in ribbon elements (Peirce and Detournay 2008)). The first option involves the net-pressure, which for low stress-contrasts is found quite inaccurately, especially in ribbon elements (Linkov 2019). The second option employs the opening, calculated much more accurately. It requires to express the intensity $R_0$ via the opening. From equation $K_{IB} = p_B \sqrt{\pi x_*}$ for a straight fracture of the half-length $x_*$ (e.g., Rice 1968), it follows $p_B = K_{IB}/\sqrt{\pi x_*}$. Then using equation (18) for the driving SIF, and substitution the result into (7) gives

$$R_0 = \frac{8\sqrt{2}}{3} \frac{\Delta\sigma}{E'} \frac{\sqrt{x_* d}}{w_B} \tag{36}$$

Illustrate the two options by the data of the examples, given in Chen et al. (2020) for a fracture propagating between parallel barriers. To use (7), of essence is that for water roughly $p_B = 1 \ MPa$, $w_B = 1 \ mm$. For gel, these values are three-fold greater. Besides, four values of stress contrasts, used in the cited paper were $\Delta\sigma = 1.0, 2.0, 3.0, 4.0\ MPa$. The corresponding *theoretical* values of the intensities for water were $R_0 = 1.17, 2.34, 3.50, 4.68$; for gel, they are $R_0 = 0.37, 0.74, 1.11, 1.48$.

For *water*, using $p_B = 1\ MPa$ in (7) yields $R_0 = 1.0, 2.0, 3.0, 4.0$ when $\Delta\sigma = 1.0, 2.0, 3.0, 4.0\ MPa$, respectively. These values are 17% less than the theoretical values $R_0 = 1.17, 2.34, 3.50, 4.68$. The agreement looks strikingly good. It may be a consequence of using quite coarse mesh (four elements along the half-height). Using $w_B = 1\ mm$ in (36) and the input parameters $E' = 3 \cdot 10^4\ MPa$, $x_* = d = 10\ m$ of the example, gives $R_0 = 1.26, 2.51, 3.77, 5.03$. These intensities are 8% greater than the theoretical values given above. Again, the agreement is unexpectedly good having in mind that the estimation (36) is quite rough.

For *gel*, the numerical results, given in Fig. 17 of the paper (Chen et al. 2020), show that both the net-pressure $p$ and the opening $w_B$ are three-fold greater than for a fracture driven by water. Hence, the estimations of intensities $R_0$ by using (7), (36) are now three-fold less. By (7) they are $R_0 = 0.33, 0.66, 0.99, 1.32$. By (36), $R_0 = 0.42, 0.84, 1.26, 1.68$. Again they agree with the theoretical values $R_0 = 0.37, 0.74, 1.11, 1.48$. Much greater errors up to $50 - 70\ \%$ would be acceptable to distinguish a case, when a barrier is certainly weak ($R_0 < 1$) and the propagation is nearly contrast-free, from a case, when a barrier is certainly high enough ($R_0 > 10$) to make the propagation channelized.

When an estimation shows that the stress contrast is weak, the conventional calculations do not require caution. Otherwise, it is reasonable to trace the net-pressure or/and opening histories to conclude if changes in the computational scheme are desirable.

## 4.2 Numerical indication of desirable changes in computational scheme

In the case, when the rough estimations of the previous subsection show that at the moment of reaching a barrier its intensity $R_0$ exceeds 3.5, the dominant propagation near the barrier may be along it. Now much depends on the conditions at other parts of the fracture front. For instance, if the high intensity $R_0$ occurs merely at the upper boundary of a pay-layer, while there is no stress-contrast on its lower boundary, then with growing distance from the source, the solution tends to that corresponding to the solution for contrast-free propagation ($R_0 = 0$) of the fracture driven by the source with two-fold less pumping rate (Linkov, Rybarska-Rusinek and Rejwer-Kosińska 2023). Clearly, *no changes* in the computational scheme are needed. In this case, the net-pressure decreases in time as $t^{b_P}$ where $b_P = -n/(n+2)$.

In the opposite case, when strong stress-contrasts occur at the both boundaries of a pay-layer, the propagation becomes entirely channelized within the layer. The net-pressure, as mentioned, after reaching the barriers becomes nearly uniform across the pay-layer. This meets the key suggestion of a P3D model and justifies its using. The maximal and average openings in a cross-section become proportional to the net-pressure, and the pressure and opening change in time with the same exponent $b_P$ in their monomial dependence $t^{b_P}$ (Linkov, Rybarska-Rusinek and Rejwer-Kosińska 2023). For a strong barrier, $b_P$ is positive. In the case of a Newtonian fluid, it is $b_P = 0.2$ for impermeable and $b_P = 0.125$ for a highly permeable pay-layer. The *pressure gradient decreases* as $1/t^{1-b_P}$, and the Poiseuille-type equation implies that the flux is defined by the product of two terms, one of which grows in time as $t^{3b_P}$, while the other decreases as $1/t^{1-b_P}$. Since $b_P \leq 0.2$, the maximal product is of the form $t^{0.6} \cdot \left(\frac{1}{t^{0.8}}\right)$. It decreases as $t^{-0.2}$ in time. Therefore, no computational problems, caused by uncertainty of the type $\infty \cdot 0$, arises. Still, problems may arise due to another reason, discussed in Subsection 2.2. Specifically, when modeling a channelized propagation with using a spatial grid, the number of unknowns grows in time proportionally to the fracture length; for a Newtonian fluid, it is proportional to $t^{0.8}$. Then in the time interval from seconds to first hours, the number of unknowns grows nearly three orders. This drastically increases the number of work units, the cost of a single work unit (single matrix-to-vector multiplication) and also the condition number of the elasticity matrix. The practical implication is: in the case, when the propagation tends to become channelized ($R_0 > 3.5$), it is reasonable to change a truly 3D computational scheme to much more robust, stable and accurate 1D scheme of the *P3D model* in its improved form (Linkov and Markov 2020). The latter may be notably simplified by using the asymptotic approach in the form (21) to simplify tracing the fracture height, since the improved P3D suffices using simple analytical formulae to calculate SIFs.

At last, in the case, when a strong barrier has a closed contour surrounding the injection source, both the average net-pressure and average opening linearly grow in time. As noted, this implies that with growing time, computations, even performed with double precision, would deteriorate. The implication is: it is reasonable to change the computational method when modeling a time interval of practical interest. In this case, the problem is promptly removed by employing the *asymptotic approach*. It is sufficient to use (21) and to solve a linear elasticity problem for a planar crack under uniform pressure.

In brief, the distinct differences between the cases considered are characterized by the exponent $b_P$ in the monomial approximation of the calculated or measured net-pressure as function of time. Actually, $b_P$ is the Nolte-Smith parameter (Nolte and Smith 1981; Nolte 1989), defined as the slope of the net-pressure history in the log-log scale. Its worth for identification of the stress-contrasts has been studied in the paper (Linkov, Rybarska-Rusinek and Rejwer-Kosińska 2023). The present study shows its significance for a choice of a proper computational scheme. For nearly *contrast-free* propagation, the Nolte-Smith slope parameter $b_P$ is negative; the net-pressure decreases in time. There is no need to change a computational scheme. When the propagation is *channelized*, the parameter does not exceed 0.2 ($0.125 < b_P \leq 0.2$); the net-pressure slowly grows in time. Then it is reasonable to change the scheme to that of the improved P3D model. In the case, when the propagation is *entirely hampered* by the barrier, $b_P = 1$, the net-pressure linearly grows in time. Then by (21), it is necessary to use the asymptotic approach in frames of the linear elasticity theory.

## 5 Summary

The paper presents the theory of hydraulic fracture propagation through a stress barrier. It reveals when and why the conventional schemes of spatial discretization may become inefficient. On this basis, we give simple recommendations how to indicate and overcome computational difficulties caused by a barrier.

Specifically, the results include the following items.

(i) The *intensity* (strength) of a stress barrier is characterized by the ratio of the stress-contrast and net-pressure near a barrier. It may change from zero (for contrast-free propagation) to infinity (for channelized propagation described by the PKN model).

(ii) The *general* computational difficulties, arising when the Courant time becomes small, while the condition number grows for fine spatial grids, are aggravated by a barrier. The complications are caused by the growth of the fracture opening $w$, which strongly (as $1/w^3$) reduces the Courant time.

(iii) A *specific* difficulty arises for high barriers along the entire fracture front. In this case, the Poiseuille-type equation involves the uncertainty $\infty \cdot 0$ in the product of a positive degree of (growing) opening by the (decreasing) pressure gradient, calculated to find the flux. Then, with growing interval of modeling, rounding errors lead to unacceptable errors of the product, and any scheme, employing spatial discretization, fails with time growth.

(iv) The difficulties are removed by the *asymptotic approach*, which avoids spatial discretization and explicit using the Poiseuille-type equation. The approach employs that almost constant net-pressure near a strong barrier yields square-root asymptotics of stresses and opening. Then SIFs may serve to formulate the propagation condition in frames of linear fracture mechanics. In general, viscous resistance is taken into account by means of the correspondence principle via the apparent viscous SIF. Finally, the penetration into a barrier is described asymptotically by the ODE in terms of the SIFs only. Commonly, the input of viscous resistance is negligible, and the equation becomes algebraic with simple analytical solution.

(v) The asymptotic approach is used to formulate the *bench-mark* problems (plain-strain and axisymmetric) for barriers of arbitrary intensity. The formulation is given in properly normalized variables, so that the number of input parameters entering the resulting Cauchy problem is reduced from eight to three, at most, without changing the time scale. The *universal solution* to the problem is given. Its analysis distinguishes three typical stages of a fracture penetration into a barrier. Their quantitative characteristics are studied, and the theoretical values of the *Nolte-Smith* slope *parameter* and of the *arrest time* are obtained as functions of the barrier intensity. The accurate solutions, given for the bench-mark plane-strain and axisymmetric problems, may serve to test computational methods for problems involving stress barriers of high intensities.

(vi) Special analysis establishes the *accuracy* and *bounds* of the asymptotic approach. It appears that the approach provides physically significant and accurate results for fracture penetration into *high*, *intermediate* and even *weak* stress barriers.

(vii) On this basis, simple *practical recommendations* are given for modeling hydraulic fractures in rocks with stress barriers. Specifically, in cases, when the area of perspective HF propagation contains stress-contrasts, it is reasonable to complement conventional modeling with the control of the intensity $R_0$ of a stress barrier at the moment of its reaching, and the need in changing the computational algorithm used. The first control may be performed by using the net-pressure or opening, calculated at cells adjacent to tip elements intersected by the stress barrier. If the calculated intensity $R_0$ is less than 1.0, the barrier is certainly weak, and there is no need in its further analysis.

Otherwise, the second control is recommended to detect if changes of the computational scheme are advisable. The control may be performed by using the Nolte-Smith slope parameter $b_P$. If it is less than 0.05, the computational scheme does not require changes. When $0.05 < b_P \leq 0.125$, the scheme is acceptable, while replacing it with 1D scheme of the improved P3D model complemented with using the asymptotic approach for tracing the height growth will drastically reduce the time expense. If $0.125 < b_P \leq 0.2$, such a change is quite desirable to avoid computational complications. At last, when $b_P > 0.2$, and especially when the slope parameter approaches 1.0, it is *necessary* to use the asymptotic approach with the driving SIF, found by solving the elasticity problem for a fracture under uniform net-pressure. The recommendations may be promptly implemented in any program using spatial discretization to model fracture propagation.

**Declarations**

**Funding** The authors did not receive support from any organization for the submitted work.
**Conflict of Interest** The authors declare that they have no known competing fnancial interests or personal relationships that could have appeared to infuence the work reported in this paper.